\documentclass[a4wide,11pt]{article}
\usepackage{a4wide}

\usepackage[hidelinks]{hyperref}
\hypersetup{colorlinks=true,citecolor=blue,linkcolor=blue,
    filecolor=blue,urlcolor=blue}

\usepackage{graphicx}        
\usepackage{multicol}        
\usepackage[bottom]{footmisc}
\usepackage[lofdepth,lotdepth]{subfig}

\usepackage{multirow}

\usepackage{enumitem}
\usepackage{bold-extra}

\usepackage{bm}

\usepackage{afterpage}

\usepackage{soul}

\usepackage{rotating}

\usepackage[shortcuts]{extdash}

\usepackage{import}

\usepackage{pdfpages}

\usepackage{graphicx}

\usepackage{tikz, pgf}
\usetikzlibrary{arrows}
\usetikzlibrary{patterns}
\usetikzlibrary{plotmarks}
\usetikzlibrary{calc}

\usepackage{amsmath}
\usepackage{listings}
\newcommand{\cmark}{yes}%
\newcommand{\xmark}{no}%

\newcommand{\acfixed}{\textsc{ac\_fixed}}

\newcommand{\acfloat}{\textsc{ac\_float}}
\newcommand{\ctfloat}{\textsc{ct\_float}}
\newcommand{\flopoco}{\textsc{FloPoCo}}
\newcommand{\vfloat}{\textsc{VFLOAT}}

\usepackage{algorithm}
\usepackage{algorithmicx}
\usepackage{algpseudocode}

\algdef{SE}[DOWHILE]{Do}{doWhile}{\algorithmicdo}[1]{\algorithmicwhile\ #1}%

\begin{document}

\title{Customizing number representation and precision}
\author{Olivier Sentieys$^1$ and Daniel Menard$^2$\\
~\\
$^1$Univ. Rennes, Inria, IRISA, Rennes, Olivier.Sentieys@inria.fr\\
$^2$INSA, IETR, Rennes Daniel.Menard@insa-rennes.fr\\
}
%
%
\date{}

\maketitle

\abstract{
There is a growing interest in the use of reduced-precision arithmetic, exacerbated by the recent interest in artificial intelligence, especially with deep learning. Most architectures already provide reduced-precision capabilities (e.g., 8-bit integer, 16-bit floating point). In the context of FPGAs, any number format and bit-width can even be considered.
In computer arithmetic, the representation of real numbers is a major issue. Fixed-point (FxP) and floating-point (FlP) are the main options to represent reals, both with their advantages and drawbacks. 
This chapter presents both FxP and FlP number representations, and draws a fair a comparison between their cost, performance and energy, as well as their impact on accuracy during computations.
It is shown that the choice between FxP and FlP is not obvious and strongly depends on the application considered. In some cases, low-precision floating-point arithmetic can be the most effective and provides some benefits over the classical fixed-point choice for energy-constrained applications. 
}

\tableofcontents

\section{Introduction }
\label{sec:intro}
There is a growing interest in the use of reduced-precision arithmetic, exacerbated by the recent interest in artificial intelligence, especially with deep learning. CPU, GPU and TPU architectures already provide interesting, but limited, reduced-precision capabilities. 8-bit integer, 16-bit floating point (e.g., \texttt{float16}, \texttt{bfloat16}) are typical examples of low-precision computations included in the architectures. Through the use of hardware acceleration on FPGA architectures, and thanks to their reconfiguration features, arithmetic customisation can be further extended and almost any number format and word-length can be leveraged in the accelerator.
All these examples illustrate the growing interest in the use of custom arithmetic. 

In computer arithmetic, the representation of real numbers is a major issue. Indeed, most algorithms are using mathematical functions, and their accuracy and stability is directly related to the accuracy of the number representation they use. 
To represent real numbers, there exist two main formats: fixed-point and floating-point. Fixed-point (FxP) representation encodes real numbers as an integer value scaled by a fixed factor, thus leading to a format comprising an integer part and a fractional part, the point of the real number being at a fixed position. In the floating-point (FlP) representation, the scaling factor is encoded in the format, which comprises a mantissa (or significand) and an exponent, the point being floating along with the computations. 

Fixed-point arithmetic is sometimes favoured due to its high efficiency in terms of energy consumption, cost, and performance, with a reputed clear advantage compared to floating-point. This comes at the cost of the pain of the programmer, who needs to manage all scaling operations to respect the rules imposed by FxP arithmetic.
Floating-point representation can be considered as the main representation for real numbers, especially in high-performance computing. In contrast to FxP,
FlP provides a high dynamic range, is able to represent with high accuracy both small and large numbers, and is very easy from a programmer point of view, since all scaling and rounding operations are totally managed by the hardware. However, this ease of use comes with relatively important area, delay and energy penalties when compared to FxP.

This chapter presents both number representations, and tries to draw a fair a comparison between customized fixed-point and floating-point arithmetic.
One conclusion is that the choice between FxP and FlP is not obvious and depends on the application considered. It is shown that, in some cases, low-precision floating-point arithmetic can be the most effective and provides some benefits over the classical fixed-point choice for energy-constrained applications. Indeed, combining the ease of use of floating-point representation associated to low-energy benefits of small bit-width, make reduced-precision floating-point arithmetic very promising, but not always useful. \\

Section \ref{sect: Fixed-point arithmetic} presents in detail the fixed-point representation, the rules governing the propagation of the fixed-point formats through operations, the quantization error process associated with computations relying on reduced-precision fixed-point arithmetic, and how overflow should also be considered. Overflow is critical in FxP since the dynamic range to represent real values is very limited in this format.

As already mentioned, a fixed-point number is composed of an integer and a fractional part. The aim of the fixed-point conversion process is to determine for each data the binary-point position and more specifically the number of bits for the integer part and the fractional part. This process is explained in Section \ref{sec:Fixed-point conversion process}, more details can also be found in Chapter 9.

Section \ref{sect: Floating-point arithmetic} details the floating-point representation, the principle of FlP addition and multiplication, and provides some fair comparisons of their cost and performance with regard to FxP. Section \ref{sect: Floating-point arithmetic} also presents some opportunities to reduce the cost of FlP operators as well as some libraries that can be used to simulate and perform hardware synthesis of customized, low-precision floating-point computations.

Finally, Section \ref{section:fxp_vs_flt} gives some comparison results in terms of area, delay, and energy between the two number representations FxP and FlP, first at the operator level, and then in the context of their use in applications, thus considering the errors due to low-precision computations.

\section{Fixed-point arithmetic}
\label{sect: Fixed-point arithmetic}

\subsection{Fixed-point representation}
\label{sec:FxP_representation}

\begin{figure}[h]
\centering
\includegraphics[width=0.6\linewidth]{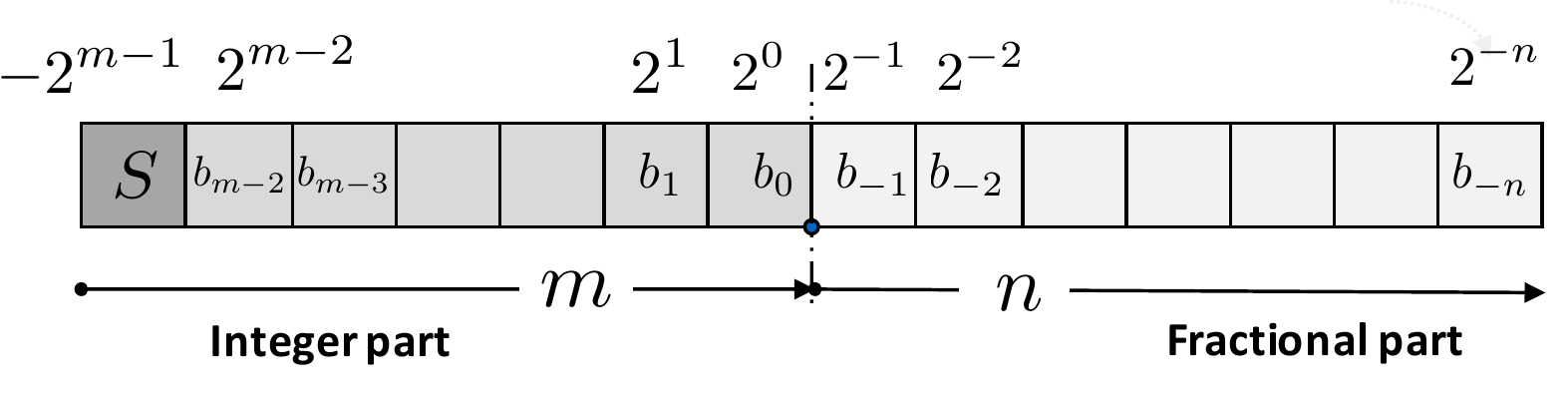}
\caption{Fixed-point specification }
\label{Fig: fxpt_specif}
\end{figure}

Fixed-point (FxP) representation is a way to encode real numbers with a virtual binary-point (BP) located between two bit locations as shown  in Figure \ref{Fig: fxpt_specif}. A fixed-point number is made-up of an integer part (left to the BP) and a fractional part (right to BP). The term $m$ designates the integer part word-length (IWL) and corresponds to the number of bits for the integer part when this term is positive. This IWL includes the sign bit for signed numbers. The term $n$ designates the fractional part word-length (FWL) and corresponds to the number of bits for the fractional part when this term is positive. The fixed-point value $x_{fxpt}$  is computed from the following relation 
\begin{equation}
x_{fxpt} = -2^{m-1}.S + \sum_{i=-n}^{m-2} b_i 2^i 
\end{equation}
Numbers in the dynamic range $[-2^{m-1}, 2^{m-1}-2^{-n}]$ can be represented in this fixed-point format with a precision of $q=2^{-n}$. The term $q$ corresponds to the quantisation step and is equal to the weight of the least significant bit $b_{-n}$. 
The \emph{Q-format} notation can be used to specify fixed-point numbers. For a fixed-point number having an IWL and FWL equal to $m$ and $n$, respectively, the notation $Q_{m.n}$ is used for signed numbers and $uQ_{m.n}$ for unsigned numbers. 
The total number of bits $w$ is equal to $m+n$. In  fixed-point arithmetic, $m$ and $n$ are fixed and lead to an implicit scaling factor equal to $2^{-n}$ which does not change during the processing. The fixed-point value $x_{fxpt}$ of the data $x$ can be computed from the integer value $x_{int}$ of the data $x$ such as %
$x_{fxpt} = x_{int} . 2^{-n}   $

\subsection{Format propagation}\label{sec: format propagation}
\label{sec:format_propagation}

In this section the rules governing the propagation of the fixed-point formats through operations are described for the different arithmetic operations. Let us consider an operation $\diamond$ having $x$ and $y$ as input operands and $z$ as output operand. Let $Q_{m_x.n_x}$, $Q_{m_y.n_y}$ and $Q_{m_z.n_z}$ be the \emph{Q-format} of the operand $x$, $y$ and $z$, respectively.  

\paragraph{Addition - subtraction}  The addition or the subtraction of two fixed-point numbers $x$ and $y$ can lead to an overflow if the operation result is not in the dynamic range of $x$ and $y$. In this case one more bit must be used to represent the integer part. Thus, dynamic range of the output result must be taken into account. A common IWL, $m_{c}$, must be defined to  represent the input and the output 

\begin{equation}
 m_{c} =  \max(m_x, m_y, m_z)  
\end{equation}
where $m_z$ is computed from the dynamic range of the variable $z$. This  IWL allows aligning the binary-point of the two input operands before computing the addition or the subtraction. The fixed-point format of the operation output is as follow

\begin{equation}
\left\{
\begin{array}{rcl}
 m_{z} &=&  m_c  \\
 n_{z} &=&  \max(n_x, n_y)
\end{array}\right.
\end{equation}

\paragraph{Multiplication} In contrast to the addition or the subtraction, there is no risk of overflow for the multiplication if the format of the output respects the following conditions. Thus, the fixed-point format of the output $z = x \times y$ is obtained from the input $x$ and $y$ fixed-point format with the following expression  

\begin{equation}
\left\{
\begin{array}{rcl}
 m_z &=&  m_x + m_y  \\
 n_z &=&  n_x + n_y
\end{array}\right. 
\end{equation}

The multiplication leads to an increase of the number of bits to represent the operation output. The total number of bits $w_z$ is equal to $w_x + w_y = m_x + n_x + m_y + n_y$.    

\paragraph{Division} For the division operation $z = x / y$, the value 0, must be excluded of the divisor $y$ interval $[\underline{y}, \overline{y}]$ leading to the interval $[\underline{y},-2^{-n_y}] \cup [2^{-n_y}, \overline{y}]$ if we consider the case that $\underline{y}$ is strictly negative and $\overline{y}$ is strictly positive. The IWL of the division output must be able to represent the largest  value of the division result. This one is obtained by dividing the largest dividend by the smallest divisor. The largest possible dividend is $-2^{m_x-1}$ while the smallest divisor is $2^{- n_y}$. 

The FWL of the division output must be able to represent the smallest absolute value of the division result. This one is obtained by dividing the smallest dividend by the largest divisor. The smallest dividend is $2^{-n_x}$ while the largest divisor is $-2^{m_y-1}$.  

Thus, the fixed-point format of the output $z = x / y$ is obtained from the input $x$ and $y$ fixed-point format with the following expression  
\begin{equation}
\left\{
\begin{array}{rcl}
 m_z &=&  m_x+n_y  \\
 n_z &=&  n_x+m_y
\end{array}.
\right.
\end{equation}
Like for the multiplication, the total number of bits $w_z$ is equal to $w_x + w_y = m_x + n_x + m_y + n_y$. 

\subsection{Quantisation process and rounding modes}
\label{sec:quantize}
In DSP applications, a sequence of arithmetic operations leads to an increase of data word-length when multiplication and division operations are involved. To maintain data word-lengths in reasonable range, the number of bits must be reduced. In fixed-point arithmetic, the least significant bits are discarded.  
Let $x'$, be a fixed-point variable with a word-length of $w_{x'}$ bits. The quantisation process $Q()$ leads to the variable $x$, depicted in Figure \ref{Fig: fxpt_specif}, and having a word-length $w = w_{x'} -d$. Let $S_x$ be the set containing all the values which can be represented in the format after quantisation. 

\paragraph{Truncation}

In the case of truncation, the data $x$ is always rounded towards the lower value available in the set $S_x$:
\begin{equation}\label{Eq: T 0}
    x  = \lfloor x \cdot q^{-1} \rfloor \cdot  q = 
    \begin{array}{ll}
       k    q  &  \forall x \in [k \cdot q; (k+1)q[ \\ 
    \end{array}
\end{equation}
with $\left\lfloor \cdot \right\rfloor$, the floor function defined as
$\left\lfloor x \right\rfloor = \max\left(n \in \mathbf{Z}
\right | n \leq x)$ and $q=2^{-n}$ the quantisation step.  
The value $x$ after quantisation is always lower or equal to the value $x$ before quantisation. Thus, the truncation adds a bias on the quantised signal and the output quantisation error will have a non zero mean. Truncation rounding is widely used because of its cheapest implementation. The $d$ LSB of $x'$ are discarded and no supplementary operation is required.

\paragraph{Conventional rounding}

To improve the precision after the quantisation, the rounding quantisation mode can be used. The latter significantly decreases the bias associated with the truncation. This quantisation mode rounds  the value $x$ to the nearest value available in the set $S_x$:
\begin{equation}\label{Eq: R conventional 1}
    x  = \left\lfloor \left(x + \frac{1}{2}q \right)  \cdot q^{-1} \right\rfloor  \cdot q =
    \left\{
    \begin{array}{ll}
       k  q  &  \forall x \in [k \cdot q;(k+ \frac{1}{2})q [ \\
      (k+1) q  &  \forall x \in [(k+ \frac{1}{2}) q ; (k+1)q] \\ 
    \end{array} \right.
\end{equation}
The midpoint $q_{1/2} = (k+ \frac{1}{2})q$ between $k q$ and
$(k+1)q$ is always rounded up to the higher value $(k+1) q$. Thus, the distribution of the quantisation error is not exactly symmetrical and a small bias is still present.

The conventional rounding can be directly implemented from (\ref{Eq: R conventional 1}). The value  $2^{-n-1}$ is added to $x'$ and then the result is truncated on $w$ bits. In the technique presented in \cite{Lapsley96_BDTI}, the conventional rounding is obtained by the addition of $x'$ and the value $b_{-n-1}.2^{-n}$ and then the result is truncated on $w$ bits. This implementation requires an adder of $w$ bits.

\paragraph{Convergent rounding}

To reduce the small bias associated with the conventional rounding, the convergent rounding can be used. To obtain a
symmetrical quantisation error, the specific value $q_{1/2}$ must be rounded-up to $(k+1)q$ and rounded-down to $k q$ with the same probability. The probabilities that a particular bit is 0 or 1 are assumed to be identical  and thus the rounding direction can depend on the bit $b_{-n}$ value.
\begin{equation}\label{Eq:conv rouning}
    x=\left\{
    \begin{array}{ll}
       k    q  &  \forall \ x \in [k.q;(k+ \frac{1}{2})q [ \\
      (k+1) q  &  \forall \ x \in ](k+ \frac{1}{2}) q ; (k+1)q] \\ 
       k    q  &  \forall \ x = q_{1/2} \quad \mathrm{and} \quad b_{-n} = 0\\ 
      (k+1) q  &  \forall \ x = q_{1/2} \quad \mathrm{and} \quad b_{-n} = 1\\  
    \end{array} \right.
\end{equation}
The specific value $q_{1/2}$ has to be detected to modify the computation in this case. For this specific value, the addition of the data $x$ with the value $2^{-n-1}$ has to be done only if the bit $b_{-n}$ is equal to one.

The alternative to this conditional addition is to add the value $b_{-n-1}.2^{-n}$ in every case. Then, for the specific value $q_{1/2}$, the least significant bit $b_{-n}$ of the data $x$ is forced to 0 to obtain an even value. This last operation does not modify the result when $b_{-n}$ is equal to 1 and discard the previous addition operation if $b_{-n}$ is equal to 0. The convergent rounding requires a supplementary addition operation and an operation (DTC) to detect the value $2^{-n-1}$ and then to force bit $b_{-n}$ to zero.

\subsection{Overflow modes}
\label{sec:overflow}
In DSP applications, numerous processing kernels involve summations requiring to accumulate intermediate results. Consequently, the dynamic range of the accumulation variable grows and can exceed the bounds of the values that can be represented leading to overflows. When an overflow occur, if no supplementary hardware is used, the wrap-around overflow mode is considered. For the wrap-around overflow mode, the value $x_{wa}$ of variable $x$ coded with $m$ bits for the IWL is equal to 
\begin{equation}\label{Eq:overflow}
    x_{wa} = \left( ( x + 2^{m-1}) \ \mathrm{mod} \ 2^{m} \right)  - 2^{m-1}
\end{equation}
with $ \mathrm{mod}$ the modulo operation. To avoid overflow, the fixed-point conversion process described in the rest of this chapter must be followed conscientiously. Especially, the dynamic range of the different variables must be carefully evaluated for sizing the IWL. For variables having a long tail for its probability density function, the IWL can be large and thus leading to an over-estimation for numerous values. In this case, saturation arithmetic can be used to reduce the IWL. Let consider a variable $x$ coded with $m$ bits for the IWL. In saturation arithmetic, when the value $x$  is lower than $-2^{m-1}$ the value $x$ is set to $-2^{m-1}$. When the value $x$  is higher than $-2^{m-1}$ the value $x$ is set to $ -2^{m-1}-2^{-n}. $

\section{Fixed-point conversion process} 
\label{sec:Fixed-point conversion process}
As described in Section \ref{sect: Fixed-point arithmetic}, a fixed-point number is made-up of an integer part and a fractional part. The aim of the fixed-point conversion process is to determine for each data  
the binary-point position and more specifically the number of bits for the integer part and the fractional part.  

The total number of bits $w_i=m_i+n_i$ to encode a data influence the implementation cost $C$. The implementation cost reduction imply to minimize the integer and fractional part word-lengths. The reduction of the number of bit leads to unavoidable error between  the finite precision values and the infinite precision ones and thus degrades the quality of the application output. Consequently, the implementation cost minimization through word-length optimization is achieved with the constraint that the output quality degradation $\Delta \lambda$ is limited and below a maximal value $\lambda_{\max}$. This fixed-point conversion process can be modeled by the following optimization process 

\begin{equation}\label{eq: optimWL}
 \underset{\mathbf{w}}{\min} \left( C(\mathbf{w}) \right) \quad \text{subject to} \quad  \Delta \lambda(\mathbf{w}) \leq \lambda_{\max}
\end{equation}
where $\mathbf{w}$ is a $N$-length vector containing the word-lengths of the $N$ data inside the application, $C(\cdot)$ is an implementation cost function that models the cost such as area or energy consumption according to the data word-lengths. $\Delta \lambda (\cdot)$ computes the quality degradation due to the tested word-length configuration $\mathbf{w}$~\cite{Menard04_EUSIPCO,Rocher04_ICASSP,Menard08_TCSI}, and $\lambda_{\max}$ represents the maximal quality degradation tolerable by the application.

Reducing the number of bits for the integer part or for the fractional part leads to different effects. The integer part word-length $m$, defines the range of values that can be represented. When $m$ is too low, overflows occur, leading to non-linearity in the processing and a significant amplitude for the error compared to infinite precision.  As long as $m$ is higher than $m_{\min}$, the minimal value ensuring no overflow, modifying $m$ will not have effect on the output quality. When $m$ is lower than $m_{\min}$, overflow occurs and quickly the output quality is highly degraded. To determine the integer word-length, the data dynamic range is evaluated and the minimal value of $m$ ensuring no overflow or a sufficiently low overflow probability is selected.

The fractional part word-length $n$ defines the accuracy. The larger $n$ is, the smaller the error between finite and infinite precision is and the higher the accuracy is improved. Unlike for the integer part, reducing the number of bit the fractional part will progressively reduce the accuracy and increase the output quality degradation. Thus, the determination of the fractional part word-length is a trade-off between the implementation cost and the  quality degradation. This trade-off is explored through the solving of the optimization process described in Equation \ref{eq: optimWL}. The word-length of each data is optimized through the minimization of the implementation cost under quality degradation constraint. This optimization process requires three elements, an optimization algorithm, a cost function $C(\cdot)$ and a quality degradation function $\Delta \lambda (\cdot)$. The quality degradation function depends on the fractional part word-length of each data. The cost function $C(\cdot)$ requires the knowledge of the total word-length of each data. Thus, for this optimization process, the integer part word-length has to be known. 

Consequently, the fixed-point conversion process is split into two parts. Firstly, the integer part word-length is determined from the results of the data dynamic range evaluation.Secondly, the fractional part word-length is optimized by solving the optimization process described in Equation \ref{eq: optimWL}.

\subsection{Integer-Part Word-Length Determination}
\label{sec:iwl} 

The first stage of the fixed-point conversion process aims at determining the number of bits for the integer part of each data in the considered application. The goal is to minimise the number of bits while protecting against overflows which degrade significantly the application quality. Firstly, the dynamic range of each signal is evaluated. The different types of techniques available to estimate the dynamic range are presented in Section \ref{sec: dynamic range evaluation}.  Secondly, the IWL is determined from the dynamic range and the fixed-point format propagation rules. Scaling operations are inserted to adapt fixed-point formats. This process is described in Section \ref{sec: binary-position determination}.

\subsubsection{Dynamic range evaluation}\label{sec: dynamic range evaluation}	


The determination of the number of bits for the integer part requires to evaluate the signal dynamic range. existing techniques to evaluate the dynamic range can  be classified according to the targeted applications. Critical systems do not tolerate high computational errors. Any overflow occurrence may lead to a system failure or a serious quality degradation. For example, in 1996, the first launch of the Ariane 5 rocket ended in explosion due to software failure. This failure was caused by the overflow of the variable representing the rocket acceleration. 
Thus, for critical systems, the integer part word-length has to cover the entire range of possible values. In this case, the bounds should be determined by techniques that guarantee the absence of overflow occurrence and allowing to certify the data dynamic range. Techniques based on interval arithmetic or affine arithmetic satisfy these constraints, but at the expense of an overestimation of the  bounds. Statistical approaches that determine bounds from a set of simulation results can reduce the overestimation, but can not ensure the absence of overflow  occurrence. 

Overflows occur when the number of bits of the integer part is not sufficient. Overflow occurrence degrades the result quality at the system output. However, the hardware implementation cost is unnecessarily increased if the number of bits exceeds the needs. Many systems are tolerant to overflows if the probability of overflow occurrence is low enough. In this case, determining the number of bits of the integer part is a trade-off between the implementation cost and the output system quality degradation. This is translated into an optimisation problem where the integer word-length of each variable of the system is reduced while maintaining an overflow probability lower than the accepted probability \cite{Nehmeh15_JSPS}. The challenge is to estimate the probability density function (PDF) of the data in order to be able to compute the overflow probability. Stochastic approaches, which model the variable PDF by propagating data PDF model from the inputs to the system output, can be considered. 

\subsubsection{IWL determination and insertion of scaling operations}
\label{sec: binary-position determination}		

The IWL is determined by propagating the IWL thought the operations from the inputs to the outputs with the help of the dynamic range evaluation results. This propagation process uses the format propagation  rules provided in Section \ref{sec: format propagation}. 

\begin{figure}[h]
\centering
\includegraphics[width=0.4\linewidth]{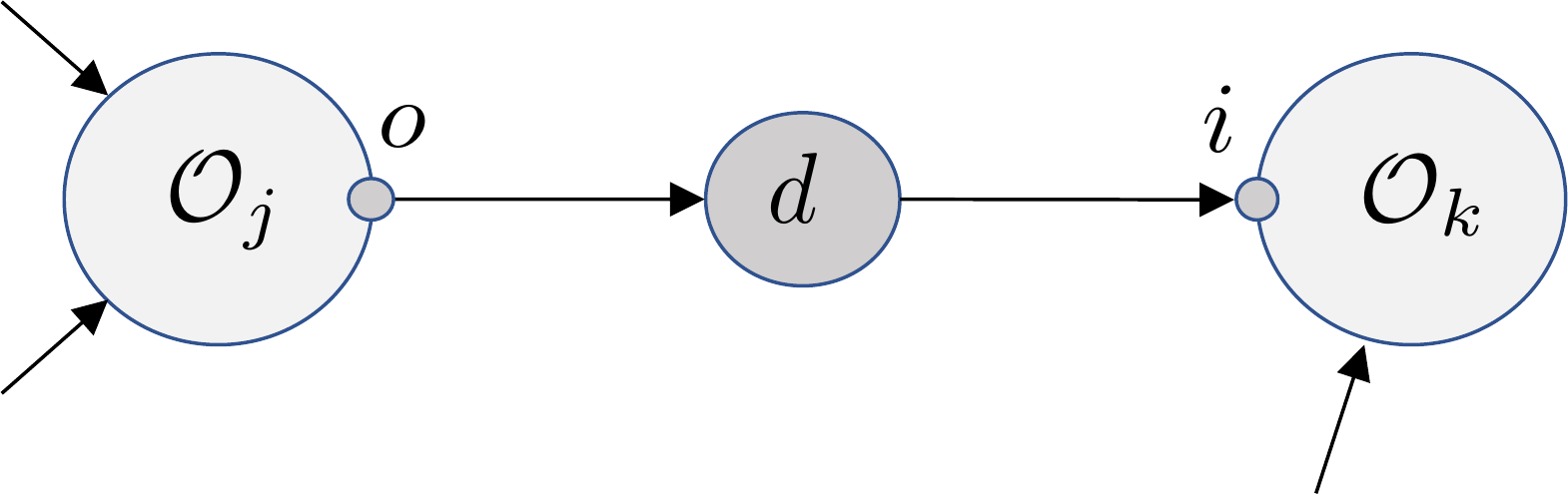}
\caption{Example of a sequence of operations: data $d$ is the output of the operation $\mathcal{O}_j$ and the input of operation $\mathcal{O}_k$.}
\label{Fig: scaling}
\end{figure}

To illustrate this stage, let us consider the sequence of operations depicted in Figure~\ref{Fig: scaling}. The data $d$ is the output of the operation $\mathcal{O}_j$ and the input of operation $\mathcal{O}_k$. Let $m_s$, $m_d$ and $m_i$ be respectively the IWL for the operation $\mathcal{O}_j$ output, the data $d$ and the operation $\mathcal{O}_k$ input.

The IWL of the signed data $d$ is computed from its dynamic range $[\underline{x}, \overline{x}]$ with the following expression   
 \begin{equation}\label{Eq: IWl computation}
 m_{x}=\max 
 \left( \lfloor \log_2(\mid \overline{x} \mid) \rfloor + 2, 
        \lceil  \log_2(\mid \underline{x} \mid)\rceil + 1
 \right).
\end{equation}
The IWL $m_o$ is computed from the propagation of the IWL of the operation $\mathcal{O}_j$ inputs with the help of the rules provided in Section \ref{sec: format propagation}. A scaling operation is inserted at the operation $\mathcal{O}_j$ output if  $s_o = m_{d} - m_{o_j}$ is strictly negative. In this case, a left shift of $s_o$ bits is required to modify the IWL. It means that the IWL of the operation $\mathcal{O}_j$ was too important compared to the data dynamic range of $d$ and the $s_o$ most significant bits of $x$ are a copy of the sign bit and can be discarded.  

For multiplication and division, the IWL $m_i$ is equal to $m_d$ and no scaling operation is required at the operation $\mathcal{O}_j$ input. For addition and subtraction, a common IWL $m_c$ for the input and the output must be determined and $m_i = m_c$.  A scaling operation is inserted at the operation $\mathcal{O}_k$ input if $s_i = m_{o_i} - m_{d}$ is strictly positive. In this case, a right shift of $s_i$ bits is required to modify the IWL of the data $d$. It means that supplementary bits are required for the addition or subtraction $\mathcal{O}_j$ to avoid overflow.

\subsection{Fractional-Part Word-Length Determination}
\label{sec:fwl}

The fractional part word-length is optimized by solving the optimization process described in Equation \ref{eq: optimWL}. The search space for this combinatorial optimization problem is huge and numerous algorithm have been proposed to find an optimized solution in a reasonable execution time. In Chapter 9 (Word-Length Optimization of Fixed-Point Algorithms), a survey of the different optimization algorithms is proposed. These optimization algorithms are iterative process, testing different combinations of word-length and moving in the search space. Consequently, the  cost function $C(\cdot)$ and the quality degradation function $\Delta \lambda (\cdot)$ are evaluated numerous time. The challenge is to develop techniques able to evaluate efficiently and accurately the quality degradation. In Chapter 9, a survey of the different existing techniques to evaluate the quality degradation is proposed. 

\section{Floating-Point Arithmetic}
\label{sect: Floating-point arithmetic}

Floating-point (FlP) representation is today the main representation for real numbers in computing, thanks to a potentially high dynamic range and to its ease-of-use since all scaling and rounding operations are totally managed by the hardware, contrary to fixed-point arithmetic. However, this ease of use comes with relatively important area, delay and energy penalties.
The floating-point representation is presented in Section \ref{sec:FlP_representation}. 
Section \ref{subsection:flp_add_mul} details the principle of FlP addition and multiplication and provides some fair comparisons of their cost and performance with regard to FxP. Then, Section \ref{subsection:flp_relax_acc} presents some opportunities to reduce the cost of FlP operators, without jeopardizing the accuracy of the computations too much. Finally, Section \ref{subsection:flp_lib} describes some libraries that can be used to simulate and perform hardware synthesis of customized, low-precision floating-point computations.

\subsection{Floating-Point Representation for Real Numbers}
\label{sec:FlP_representation}
\begin{figure}[htbp]
\centering
\includegraphics[width=0.8\linewidth]{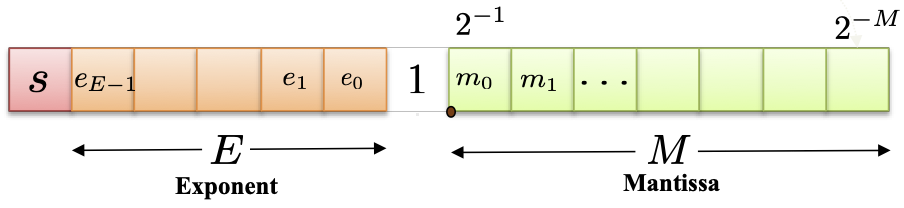} 
\caption{Floating-point representation}
\label{Fig: flpt_specif}
\end{figure}

In computer arithmetic, the representation of real numbers is a major issue. Indeed, most algorithms are using mathematical functions, and their accuracy and stability is directly related to the accuracy of the number representation they use. 
The floating-point (FlP) representation is a way to encode real numbers with a scaling factor encoded in the data. 
Given an unsigned $M$-bit mantissa $m$, a signed integer exponent of value $e$ coded on $E$ bits, often represented in biased representation, and a sign bit $s$, the radix-2 floating-point value $x_{flpt}$  is represented as 
\begin{equation}
x_{flpt} = 
	\left(-1\right)^{s} \times 1.m \times 2^{e}.
\end{equation}
Contrary to fixed-point representation, the point in the FlP representation of the number is \textit{"floating"} and scaled by the exponent, similarly to the scientific representation in decimal arithmetic that we use in our daily life. 
The mantissa $m$  -- or the significand of the representation -- is used to generate a normalized number
with an implicit "1" conforming the integer part belonging to $\left[1,2\right[$. This "1" being implicit, it is not represented in the format, freeing space for one more digit. 
This number is then scaled by means of the exponent $e$, and the sign is controlled
by the value of the sign bit $s$. 
$e$ being a signed number, the exponent is usually represented in the number format as biased by a constant value $b$.
With this representation, any number under this format can be represented using $M + E + 1$ bits as shown in Figure~\ref{Fig: flpt_specif}. 

Nevertheless, automatically keeping the \emph{floating} point at the right position along computations requires an important hardware overhead, as discussed in Section~\ref{subsection:flp_add_mul}. Managing subnormal numbers (numbers between $0$ and the smallest positive possible representable value) and the values 0 and infinity also represents an overhead. Despite this additional cost, FlP representation is today established as the \textit{defacto} standard for real number representation. Indeed, besides its high accuracy and high dynamic range, it has the huge advantage of leaving the whole management of the representation to the hardware instead of leaving it to the software designer, significantly diminishing developing and testing time. This domination is sustained by IEEE 754 standard, lastly revised in 2008~\cite{IEEESTD7542008}, which sets the conventions for floating-point number possible representation, subnormal numbers management and the different cases to be handled, ensuring a high portability of programs.
Table~\ref{tab:IEEE754nr} gives the representations of the FlP numbers following the  IEEE 754-2008 standard. Mantissa width $M$ is without the implicit $1$. The bias $b$ is equivalent to the maximum exponent value $e_{max}$.
However, such a strict normalization implies:
\begin{itemize}
\item an important overhead for throwing flags for the many special cases, and even more for the management of these special cases,
\item and a low flexibility in the widths of the mantissa and exponent, which have to respect the rules of Table~\ref{tab:IEEE754nr} for 16, 32, 64 and 128-bit precisions.
\end{itemize}

\begin{table}[htbp]
\centering
\begin{tabular}{|c|c|c|c|c|}
\hline
\multirow{2}{*}{Precision}&Mantissa&Exponent&Minimum exponent&Exponent bias ($b$)\\
&width ($M$)&width ($E$) &value ($e_{min}$) & (also $e_{max}$)\\
\hline
\hline
Half precision (16 bits)&$10$&$5$&$-14$&$15$\\
\hline
Single precision (32 bits)&$23$&$8$&$-126$&$127$\\
\hline
Double precision (64 bits)&$52$&$11$&$-1022$&$1023$\\
\hline
Quadruple precision (128 bits)&$112$&$15$&$-16382$&$16383$\\
\hline
\end{tabular}
\caption[IEEE 754 normalized floating-point representation]{IEEE 754 normalized floating-point representation}
\label{tab:IEEE754nr}
\end{table}

\subsection{Floating-Point Operators}
\label{subsection:flp_add_mul}
Integer addition (or subtraction) is the simplest arithmetic operator. However, in floating-point arithmetic, addition suffers from a high control overhead, which requires several steps to be performed:
\begin{itemize}
\item First, the difference of the exponents is computed.
\item Depending on the difference of the exponents, one among two computing paths may be selected~\cite{Muller2009HFPA}: The \textit{close path} is for situations where a massive cancellation (more than 1 bit) may occur, or effective subtractions of inputs with exponents that differ by at most 1. The \textit{far path} is for distant exponents, where their difference is at least 2 bits. The following computations may slightly vary depending on the choosen path.
\item The addition of the mantissas is performed.
\item Then, rounding is performed on the mantissa, depending on the dropped bits and the rounding mode (to zero, to nearest, etc.) selected. 
\item Special cases are then handled (zero, infinity, subnormal results), and the output sign.
\item Then, mantissa is shifted so it represents a value in $\left[1,2\right[$.
\item And the exponent is modified depending on the number of shifts.
\end{itemize}
More control can be needed, depending on the implementation of the FlP adder and the specificities of the FlP representation. For instance, management of the implicit $1$ implies to add $1$s to the mantissas before addition, and an important overhead can be dedicated to exception handling.
For a figure illustrating the FlP addition principle and more details on its hardware implementation, the reader can refer to Figure 8.13 of~\cite{Muller2009HFPA} and the related chapter of the book. \\

For cost, delay, and power comparison, Table~\ref{tab:flp_int_add_comp} shows the performance of 32-bit and 64-bit FlP addition compared with 32-bit and 64-bit integer addition, synthesized using Synopsys Design Compiler targeting 28nm FDSOI with a 200 MHz clock. Power is estimated using $10,000$ uniform input samples. FlP addition power was estimated activating in an equivalent way the \textit{close} and \textit{far} paths $50\%$ of the time.
\begin{table}[htbp]
\centering
\begin{tabular}{|c|c|c|c|c|}
\hline
~&Area&Total&Critical&Power-Delay\\
~&($\mu m^2$)&power (mW)&path (ns)&Product (fJ)\\
\hline
\hline
32-bit&\multirow{2}{*}{$653$}&\multirow{2}{*}{$4.39\mathrm{E}{-4}$}&\multirow{2}{*}{$2.42$}&\multirow{2}{*}{$1.06\mathrm{E}{-3}$}\\
float&&&&\\
\hline
64-bit&\multirow{2}{*}{$1453$}&\multirow{2}{*}{$1.12\mathrm{E}{-3}$}&\multirow{2}{*}{$4.02$}&\multirow{2}{*}{$4.50\mathrm{E}{-3}$}\\
float&&&&\\
\hline
32-bit&\multirow{2}{*}{$189$}&\multirow{2}{*}{$3.66\mathrm{E}{-5}$}&\multirow{2}{*}{$1.06$}&\multirow{2}{*}{$3.88\mathrm{E}{-5}$}\\
int&&&&\\
\hline
64-bit&\multirow{2}{*}{$373$}&\multirow{2}{*}{$7.14\mathrm{E}{-5}$}&\multirow{2}{*}{$2.10$}&\multirow{2}{*}{$1.50\mathrm{E}{-4}$}\\
int&&&&\\
\hline
\end{tabular}
\caption[Cost, delay, and power of floating-point addition vs. integer addition]{Cost, delay, and power of FlP addition vs. integer addition}
\label{tab:flp_int_add_comp}
\end{table}
These results clearly show the overhead of floating-point addition. For 32-bit, the FlP addition is $3.5\times$ larger, $2.3\times$ slower and consumes $27\times$ more energy than integer addition. For 64-bit, the FlP addition is $3.9\times$ larger, $1.9\times$ slower and consumes $30\times$ more energy. The overhead seems to be roughly linear with the size of the operator, and the impact of numbers representation is highly impacting performance. However, it is showed later in this chapter that this high difference reduces with the bitwidth of the operands. \\

FlP multiplication is less complicated than addition as only a low control overhead is necessary to perform the operation. Input mantissas are multiplied using a classical integer multiplier, while exponents are simply added. At worse, a final $+1$ on the exponent can be needed, depending on the result of the mantissas multiplication and the related rounding and normalization required. 
For a figure illustrating the basic architecture of a FlP multiplier, the reader can refer to Figure 8.14 of~\cite{Muller2009HFPA} and the related chapter of the book. 

Obviously, all classical hardware overheads needed by FlP representation are again necessary (rounding logic, normalization, management of particular cases), but the overhead is less than for addition. Table~\ref{tab:flp_int_mul_comp} shows the difference between 32-bit and 64-bit floating-point multiplication and 32-bit and 64-bit fixed-width integer multiplication, with the same experimental setup than discussed before for the addition.

\begin{table}[htbp]
\centering
\begin{tabular}{|c|c|c|c|c|}
\hline
~&Area&Total&Critical&Power-Delay\\
~&($\mu m^2$)&power (mW)&path (ns)&Product (fJ)\\
\hline
\hline
32-bit&\multirow{2}{*}{$1543$}&\multirow{2}{*}{$8.94\mathrm{E}{-4}$}&\multirow{2}{*}{$2.09$}&\multirow{2}{*}{$1.87\mathrm{E}{-3}$}\\
float&&&&\\
\hline
64-bit&\multirow{2}{*}{$6464$}&\multirow{2}{*}{$6.56\mathrm{E}{-3}$}&\multirow{2}{*}{$4.70$}&\multirow{2}{*}{$3.08\mathrm{E}{-2}$}\\
float&&&&\\
\hline
32-bit&\multirow{2}{*}{$2289$}&\multirow{2}{*}{$6.53\mathrm{E}{-5}$}&\multirow{2}{*}{$2.38$}&\multirow{2}{*}{$1.55\mathrm{E}{-4}$}\\
int&&&&\\
\hline
64-bit&\multirow{2}{*}{$8841$}&\multirow{2}{*}{$1.84\mathrm{E}{-4}$}&\multirow{2}{*}{$4.52$}&\multirow{2}{*}{$8.31\mathrm{E}{-4}$}\\
int&&&&\\
\hline
\end{tabular}
\caption[Cost, delay, and power of FlP multiplication vs. integer multiplication]{Cost, delay, and power of floating-point multiplication vs. integer multiplication}
\label{tab:flp_int_mul_comp}
\end{table}

A first observation on the area shows that the integer multiplier is $48\%$ larger than the FlP version for 32 bits, and $37\%$ larger for 64 bits. This difference is due to the smaller size of the integer multiplier in the FlP multiplier, since it is limited to the size of the mantissa (24 bits for 32-bit version and 53 bits for 64-bit version). Despite the management of the exponent, the overhead is not large enough to produce a larger operator. However, the 32-bit FlP multiplication energy is $11\times$ higher than for the integer version, while 64-bit version consumes even $37\times$ more energy. This can be justified by the higher activity of the logic in the FlP operator due the management of the exponent and special cases.
It is interesting to note that the difference of energy consumption between addition and multiplication is much more important for integer operators than for FlP. As an example, for 32-bit, integer multiplication consumes $4.7\times$ more energy than integer addition, while this factor is only $1.4\times$ for 32-bit FlP multiplier compared to 32-bit FlP adder. Therefore, using multiplication in FlP computing is relatively less penalizing than for integer multiplication, typically used in fixed-point arithmetic.

\subsection{Low-Precision Floating-Point Arithmetic}
\label{subsection:flp_relax_acc}

There is a growing interest in the use of reduced-precision arithmetic, exacerbated by the recent interest in artificial intelligence, especially with deep learning. CPU, GPU and TPU architectures already provide interesting, but limited, reduced-precision capabilities. 8-bit integer, 16-bit floating point (e.g., \texttt{float16}, \texttt{bfloat16}) are typical examples of low-precision computations included in the architectures. Through the use of hardware acceleration on FPGA architectures, and thanks to their reconfiguration features, arithmetic customisation can be further extended and almost any number format and word-length can be leveraged in the accelerator.
All these examples illustrate the interest of customizable floating-point architectures. Indeed, combining the ease of use of floating-point representation associated to low-energy benefits of small bit-width, make reduced-precision floating-point arithmetic very promising.

There are several possible opportunities to relax accuracy in floating-point arithmetic to increase performance and save power and hardware cost. Of course, the main technique is to reduce the size of the mantissa and exponent (i.e., smaller operand bit-width or word-length). With a mantissa normalized in $\left[1,2\right[$, reducing the word-length corresponds to pruning the LSBs, which comes with no overhead, except if faithful rounding is performed. For the exponent, the transformation can be more complicated if it is represented with a bias. Indeed, if $E$ is the exponent width, an implicit bias of $2^{E} - 1$ applies to the exponent in classical exponent representation. Therefore, reducing the exponent to a width $E'$ means that a new bias must be applied. The original exponent must be added $2^{E'} - 2^{E} \left(< 0\right)$ before pruning the MSBs, implying hardware overhead at conversion. The original exponent must represent a value in $\left[-2^{E'-1}+1,2^{E'-1}\right]$ to avoid overflow. In practice, it is better to keep a constant exponent width to avoid useless overhead and conversion overflows, which would have a huge impact on the quality of the computations.

A second way to improve computation at reduced precision is to play with the implicit bias of the exponent. Indeed, increasing the exponent width increases the dynamic towards infinity, but also the accuracy towards zero. Thus, if the absolute maximum values to be represented are known, the bias can be chosen so it is just large enough to represent these values. This way, the exponent gives more accuracy to very small values, increasing accuracy. However, using a custom bias means that the arithmetical operators (addition and multiplication) must consider this bias in the computation of resulting exponent, and the optimal bias along computation may diverge to $-\infty$. To avoid this, if the original $2^{E} - 1$ exponent bias is kept, exponent bias can be simulated by biasing the exponents of the inputs of each or some computations using shifting. For the addition, biasing both inputs adding $2^{E_{\text{in}}}$ to the exponent implies that the output will also be represented biased by $2^{E_{\text{in}}}$. For the multiplication, the output will be biased by $2^{E_{\text{in}}+1}$. Keeping an implicit track of the bias along computations allows to know any algorithm output bias, and to perform a final rescaling of the outputs.

Finally, accuracy can be relaxed in the integer operators composing the considered {FlP} operators, e.g., the integer adder adding the mantissas in {FlP} addition or the integer multiplier in the {FlP} multiplication. Indeed, they can be replaced by approximate adders and multipliers as described in other chapters of this book, to improve performance while relaxing accuracy. However, as most of the cost relies in control hardware, the impact on accuracy would be strong for a very small cost or performance benefit. The same approximation can be applied on the exponent management, but the impact of approximate arithmetic would be too high on the accuracy and this is therefore strongly unadvised.

\subsection{Reduced-Precision Floating-Point Libraries}
\label{subsection:flp_lib}

The past years have hosted the creation of several customizable floating-point libraries. 
As part of the synthesizable C++ libraries AC Datatypes~\cite{acdatatypes}, Mentor Graphics proposes the custom floating-point class \acfloat. Based on the fixed-point library \acfixed, \acfloat\ allows for light floating-point computation, thanks to simple operators. The mantissa in the representation is not normalized and has no implicit 1. This allows for easy management of subnormals, but induces a potential loss of accuracy in computations. The mantissa is represented in signed two's complement, so the sign information is contained in the mantissa instead of using an extra sign bit. However, there is no benefit to this choice since two's complement represents a loss of 1 bit of precision compared to unsigned representation. The choice of two's complement representation on the mantissa also turns comparison operator more complex. Moreover, many cases are not handled such as zero or infinity. \acfloat\ also supports custom exponent bias, but managing the exponent bias comes with an overhead.

\flopoco\ (for Floating-Point Cores, but not only) is a generator of arithmetic cores~\cite{DinechinIEEEDTC2011}. Also based on C++, it has its own synthesis engine and directly returns VHDL. More than simple arithmetic operators, it is able to generate optimized floating-point computing cores performing complex arithmetic expressions. In this Section, we will only get interested in \flopoco's custom floating-point addition and multiplication. The main difference of \flopoco's floating-point representation is the extra 2-bit exception field transported in data. Like for \ctfloat\, subnormals are not handled by \flopoco. Unlike \acfloat\, both \ctfloat\ and \flopoco\ do not support custom exponent bias.

Other alternatives such as \vfloat~\cite{FangACMTRTS2016,FangRTS2016} or OptiFEX~\cite{MahzoonVLSI2017} do exist but are not taken into account in the study led in this chapter. \vfloat proposes IEEE 754-2008 compliant customizable computing cores for existing {FPGA}. OptiFEX generates floating-point computing cores targeting {FPGA} like \flopoco. \\

\ctfloat\ \cite{barrois:hal-01633723}\footnote{\url{https://gitlab.inria.fr/sentieys/ctfloat}} offers a balance between computational safety and simplicity. Inspired by \acfloat, it is provided as C++ template for High-Level Synthesis (HLS), compatible with Mentor Graphics CatapultHLS and Xilinx Vivado HLS. As \ctfloat\ will be used for comparison with fixed-point representation in the rest of this chapter, we provide below more details on the library.

The declaration of an instance of \ctfloat\ requires three template parameters: the exponent width $E$, the mantissa width $M$, and
the rounding mode. 
The mantissa also includes a sign bit and is represented as sign plus absolute value, as in standard FlP. The total number of bits in memory is therefore equal to $E + M$.
Currently, two rounding modes are supported:
\texttt{CT\_RN} rounding to nearest with  halfway-to-even tie-breaking rule, and \texttt{CT\_RZ} rounding towards $0$, or truncation.
\ctfloat\ representation and arithmetic operators were created to remain simple and energy efficient, thanks to the combination of several implementation choices.
\ctfloat\ mantissa is represented in $[1,2[$ with an implicit $1$. However, subnormal numbers are not handled, which implies that a certain range of numbers are not representable around $0$.
The exponent is represented in a biased representation. The bias is set at the center of the exponent range, similar to the IEEE 754 representation. Using biased representation instead of two's-complement results in simpler exponent value comparisons, which are omnipresent in arithmetic operators. In variants of the \ctfloat\ library, the bias can also be customized.

The library provides a rich set of synthesizable operator overloading: unary operators (unary $-$, $!$, $++$, $--$), relational operators ($<$, $>$, $<=$, $>=$, $==$, $!\!=$), binary operators ($+$, $+=$, $-$ $-=$, $*$, $*\!=$, $<<$, $<<=$, $>>$, $>>=$), and assignment operator from/to another instance of \ctfloat.
It also provides non-synthesizable operator overloading features, such as conversion from/to C++ native datatypes (float, double), and output operator $<<$ for easy display and writing in files. Other built-in functions allow easy manipulation of floating-point values, such as functions to get information about the extreme representable values for a given floating-point representation, to test if a given value is representable, etc.

An example (not including all statements and declarations) of the use of \ctfloat\ is given below.
\begin{verbatim}
    ct_float<7, 9, CT_RN> h = 1.046978e-3;
    ct_float<7, 9, CT_RN> x, y;
    x = -0.02266398;
    y = x * y + 0.55;
    cout << y << endl;
\end{verbatim}
This example can be simulated and synthesized to hardware using HLS.
In the example, all variables have the same representation (i.e., $E=7$ and $M=9$, rounding mode is \texttt{CT\_RN}). It is also possible to deal with various representations.
If the inputs are on $(E_{1}, M_{1})$ and $(E_{2}, M_{2})$ representation, the output representation $(E_{o}, M_{o})$ is given by:
\begin{eqnarray}
\label{eq:ctfloat_output_determination}
E_{o} &=& \max \left(E_{1}, E_{2}\right) \nonumber \\
M_{o} &=& \max \left(M_{1}, M_{2}\right).
\end{eqnarray}
Moreover, as subnormals are not representable by \ctfloat, the output is always saturated to the smallest absolute possible representable value with the same sign. Towards infinity, the operators do not under/overflow. Saturation to the highest absolute representable value of the same sign is returned. \\

Table~\ref{tab:custom_flt_properties} recapitulates the different known properties of \acfloat, \ctfloat\ and \flopoco\ floating-point representation. In this table, the number of additional bits in the representation is taking for reference a representation with implicit $1$ in the mantissa and with one bit of sign in the representation. For an equal general accuracy, \acfloat\ needs one more bit on the mantissa than \ctfloat\ and \flopoco. However, with its 2-bit exception field, \flopoco\ has the representation requiring the largest width, but also the highest computing reliability. 
\begin{table}[htb]
\centering
\begin{tabular}{|c|c|c|c|}
\hline
 ~&\acfloat&\ctfloat&\flopoco\\
\hline
\hline
Custom exp.&\multirow{2}{*}{\cmark}&\multirow{2}{*}{\xmark (\cmark)}&\multirow{2}{*}{\xmark}\\
bias&&&\\
\hline
Mantissa&\multirow{2}{*}{\xmark}&\multirow{2}{*}{\cmark}&\multirow{2}{*}{\cmark}\\
Implicit 1&&&\\
\hline
Zero and inf.&\multirow{2}{*}{\xmark}&\multirow{2}{*}{\xmark}&\multirow{2}{*}{\cmark}\\
exception flags&&&\\
\hline
Zero and inf.&\multirow{2}{*}{\xmark}&\multirow{2}{*}{\cmark}&\multirow{2}{*}{\xmark}\\
internal handling&&&\\
\hline
Subnormal&\multirow{2}{*}{\xmark}&\multirow{2}{*}{\xmark}&\multirow{2}{*}{\cmark}\\
exception flag&&&\\
\hline
Subnormal&\multirow{2}{*}{\cmark}&\multirow{2}{*}{\xmark}&\multirow{2}{*}{\xmark}\\
internal handling&&&\\
\hline
Additional bits&\multirow{2}{*}{$+1$}&\multirow{2}{*}{$+0$}&\multirow{2}{*}{$+2$}\\
in representation&&&\\
\hline
\end{tabular}
\caption{Main properties of the custom floating-point libraries \acfloat, \ctfloat\ and \flopoco}
\label{tab:custom_flt_properties}
\end{table}

Then, the hardware performance comparison process for \acfloat, \ctfloat\ and \flopoco\ is as follows. All operators are characterized for a 28nm FDSOI @ 1.0V, 25°C ASIC library. All designs are synthesized and estimated with a clock of 200 MHz.
For power analysis, the random inputs generated for adder/subtracter characterization are ensuring an activation of the close path for at least 50\% of the computations.
However, the benchmark generated by \flopoco\ does not insure any proportion of activation of the close path, so the dynamic power could be underestimated. Moreover, \flopoco's benchmark does not consider any input and output data registers, whereas \acfloat\ and \ctfloat, synthesized with HLS, do. This may represent about 5 to 10\% underestimation in the total power for \flopoco\ operators, which has to be kept in mind for the analysis of results.
All operators are generated so they execute in 1 cycle. It may not be the most efficient implementation because of possible glitches, but it is a good starting point for a fair comparison. \\

For this comparative study, half-precision ($E=5$, $M=11$) and single-precision ($E=8$, $M=24$) floating-point representations are considered.
Results for 16-bit (resp. 32-bit) addition/subtraction (resp. multiplication) are given in Tables~\ref{tab:ctflt_16add},~\ref{tab:ctflt_32add},~\ref{tab:ctflt_16mul} and~\ref{tab:ctflt_32mul}. The two last lines of the tables refer to the relative performance of  \ctfloat\ vs. \acfloat\ (resp. \flopoco ) (e.g., \ctfloat\ area is 2.15\% higher than \acfloat).
\begin{table}[htb]
\centering
\begin{tabular}{|c||c|c|c|c|}
\hline
&\multirow{2}{*}{Area ($\mu m^2$)}&Critical&Total&Energy per\\
&&path (ns)&power (mW)&operation (pJ)\\
\hline
\hline
\acfloat & $312$ & $1.44$ & $1.84 \mathrm{E}{-1}$ & $9.07 \mathrm{E}{-1}$\\
\hline
\ctfloat & $318$ & $1.72$ & $2.13 \mathrm{E}{-1}$ & $1.05$\\ 
\hline
\flopoco & $361$ & $2.36$ & $1.84 \mathrm{E}{-1}$ & $9.06 \mathrm{E}{-1}$\\
\hline
\hline
\textbf{\ctfloat /\acfloat }& \textbf{+2.15\%} & \textbf{+19.4\%} & \textbf{+15.4\%} & \textbf{+15.7\%}\\
\hline
\textbf{\ctfloat /\flopoco} & \textbf{-11.8\%} & \textbf{-27.0\%} & \textbf{+15.7\%} & \textbf{+15.8\%}\\
\hline
\end{tabular}
\caption{Comparative results for 16-bit FlP addition/subtraction with $F_{\texttt{clk}} = 200 \text{MHz}$}
\label{tab:ctflt_16add}
\bigskip
\begin{tabular}{|c||c|c|c|c|}
\hline
&\multirow{2}{*}{Area ($\mu m^2$)}&Critical&Total&Energy per\\
&&path (ns)&power (mW)&operation (pJ)\\
\hline
\hline
\acfloat & $488$ & $1.18$ & $2.15 \mathrm{E}{-1}$ & $1.05$\\
\hline
\ctfloat & $389$ & $1.13$ & $1.76 \mathrm{E}{-1}$ & $8.59 \mathrm{E}{-1}$\\ 
\hline
\flopoco & $361$ & $1.52$ & $1.34 \mathrm{E}{-1}$ & $6.50 \mathrm{E}{-1}$\\
\hline
\hline
\textbf{\ctfloat /\acfloat }& \textbf{-20.4\%} & \textbf{-4.24\%} & \textbf{-18.2\%} & \textbf{-18.2\%}\\
\hline
\textbf{\ctfloat /\flopoco} & \textbf{+7.68\%} & \textbf{-25.6\%} & \textbf{+31.7\%} & \textbf{+32.1\%}\\
\hline
\end{tabular}
\caption{Comparative results for 16-bit FlP multiplication with $F_{\texttt{clk}} = 200 \text{MHz}$}
\label{tab:ctflt_16mul}
\end{table}
\begin{table}[htb]
\centering
\begin{tabular}{|c||c|c|c|c|}
\hline
&\multirow{2}{*}{Area ($\mu m^2$)}&Critical&Total&Energy per\\
&&path (ns)&power (mW)&operation (pJ)\\
\hline
\hline
\acfloat & $678$ & $2.49$ & $4.46 \mathrm{E}{-1}$ & $2.21$\\
\hline
\ctfloat & $720$ & $2.84$ & $4.86 \mathrm{E}{-1}$ & $2.41$\\ 
\hline
\flopoco & $772$ & $4.10$ & $5.05 \mathrm{E}{-1}$ & $2.51$\\
\hline
\hline
\textbf{\ctfloat /\acfloat }& \textbf{+6.06\%} & \textbf{+14.1\%} & \textbf{+8.92\%} & \textbf{+9.12\%}\\
\hline
\textbf{\ctfloat /\flopoco} & \textbf{-6.85\%} & \textbf{-30.8\%} & \textbf{-3.69\%} & \textbf{-4.15\%}\\
\hline
\end{tabular}
\caption{Comparative results for 32-bit FlP addition/subtraction with $F_{\texttt{clk}} = 200 \text{MHz}$}
\label{tab:ctflt_32add}
\bigskip
\begin{tabular}{|c||c|c|c|c|}
\hline
&\multirow{2}{*}{Area ($\mu m^2$)}&Critical&Total&Energy per\\
&&path (ns)&power (mW)&operation (pJ)\\
\hline
\hline
\acfloat & $1,689$ & $2.19$ & $1.02$ & $5.03$\\
\hline
\ctfloat & $1,469$ & $2.30$ & $5.84 \mathrm{E}{-1}$ & $2.70$\\ 
\hline
\flopoco & $2,890$ & $3.20$ & $1.03$ & $5.07$\\
\hline
\hline
\textbf{\ctfloat /\acfloat }& \textbf{-13.0\%} & \textbf{+5.02\%} & \textbf{-42.8\%} & \textbf{-46.3\%}\\
\hline
\textbf{\ctfloat /\flopoco} & \textbf{-49.2\%} & \textbf{-28.2\%} & \textbf{-43.3\%} & \textbf{-46.8\%}\\
\hline
\end{tabular}
\caption{Comparative results for 32-bit FlP multiplication with $F_{\texttt{clk}} = 200 \text{MHz}$}
\label{tab:ctflt_32mul}
\end{table}

The main conclusion is that the three custom floating-point libraries provide results in the same order of magnitude. For 16-bit addition/subtraction, \ctfloat\ consumes 15\% more energy than both \acfloat\ and \flopoco, despite an area being equivalent to \acfloat\ and 12\% smaller than \flopoco. The fastest 16-bit adder/subtracter is \acfloat, followed by \ctfloat, which is 19\% slower but 27\% faster than \flopoco. All metrics are slightly in favour of \acfloat\ for 16-bit addition/subtraction.
For 16-bit multiplication, \flopoco's multiplier is the smallest and with the lowest energy consumption. However, \ctfloat\ is 25\% faster but consumes 32\% more energy. 

32-bit addition/subtraction gives similar energy for \acfloat, \ctfloat\ and \flopoco. \flopoco\ is the slowest operator, \ctfloat\ being 27\% faster.
The energy of 32-bit multiplication is strongly in favour of \ctfloat, saving more than 45\% more energy than both \acfloat\ and \flopoco. \ctfloat\ is 13\% smaller than \acfloat\ and 49\% smaller than \flopoco. However, \acfloat\ is 5\% faster. 

As a conclusion, \acfloat, \ctfloat\ and \flopoco\ addition/subtraction and multiplication provide similar results. Though they all have different features (implicit 1 or not, particular cases management, etc.), they all are quite close in terms of performance. 
In the following section, \ctfloat\ alone is then used as a reference for the comparison with fixed-point arithmetic.


\section{Comparison between fixed-point and custom floating-point}
\label{section:fxp_vs_flt}

This section draws a comparison between customized fixed-point and floating-point arithmetic. Section \ref{section:fxp_vs_flt_standalone} compares FxP and FlP in terms of area, delay, and energy at the operator level. Then, Section \ref{section:fxp_vs_flt_app} compares the two number representations in the context of their use in applications, thus considering the errors due to low-precision computations. One conclusion is that the choice between FxP and FlP is not obvious and depends on the application considered. It is shown that, in some cases, low-precision floating-point arithmetic can be the most effective, providing some benefits over the classical fixed-point choice for energy-constrained applications.

\subsection{Operator-level comparison}
\label{section:fxp_vs_flt_standalone}
This section compares FxP and FlP operators in terms of area, delay, and energy, and does not consider computing errors at the operator level. Indeed, floating-point error magnitude is related to data values. Low-amplitude data have low error magnitude, whereas high amplitude data have much higher error magnitude. Oppositely, fixed-point has a very homogeneous error magnitude, uniformly distributed between fixed bounds. Therefore, its relative error depends on the amplitude of the represented data. It is low for high amplitude data and high for low amplitude data. This duality makes these two paradigms impossible to be atomically compared using the same error metric. The only interesting error comparison which can be performed is to compare the error behaviour inside the same application, which is reported in Section~\ref{section:fxp_vs_flt_app} on {FFT} and K-means clustering.

The study in Section \ref{section:fxp_vs_flt} uses the \ctfloat\ library for custom floating-point and \acfixed\ datatypes. A 100~MHz clock is set for synthesis and power estimation. All the other parameters are the same as for the previous section.

In this section, 8-, 10-, 12-, 14- and 16-bit operators are compared. For each of these bit-widths, several versions of the floating-point operators are estimated with different exponent widths and compared with fixed-point. $25 .10^{3}$ uniformly distributed inputs are used for each operator characterization. For the floating-point adder, inputs are distributed such that  the close path, which has the highest energy by nature, is activated 25\% of the time.
Adders and multipliers are all tested in their fixed-width version, meaning their number of input and output bits are the same. The output is truncated. 

\begin{figure}[htbp]
\begin{center}
\includegraphics[width=0.8\columnwidth]{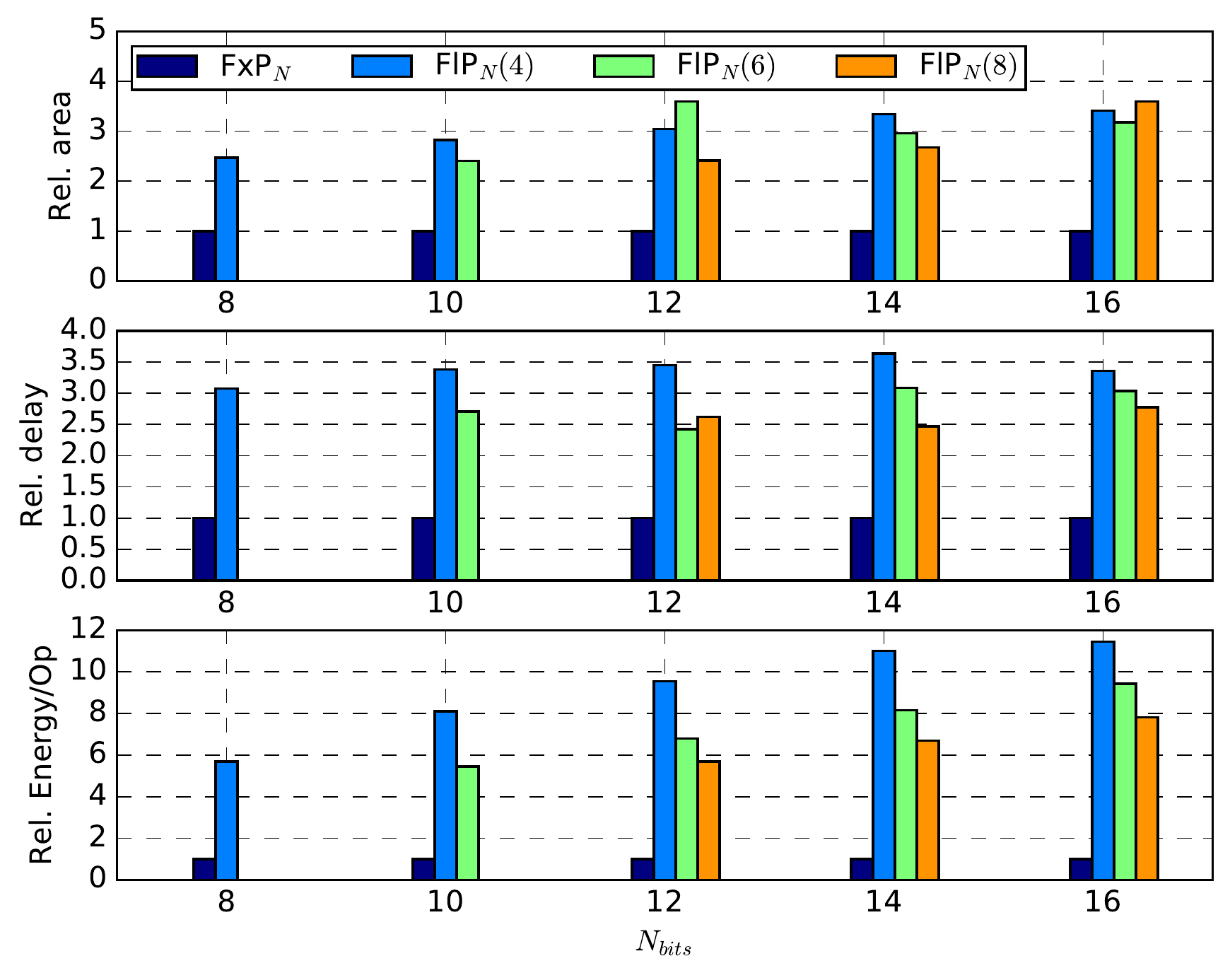}%
\caption{Relative area, delay and energy-per-operation between fixed-point and floating-point of adders for different bit-widths}
\label{fig:fxp_vs_FlP_add}
\end{center}
\end{figure}
\begin{figure}[ht]
\begin{center}
\includegraphics[width=0.8\columnwidth]{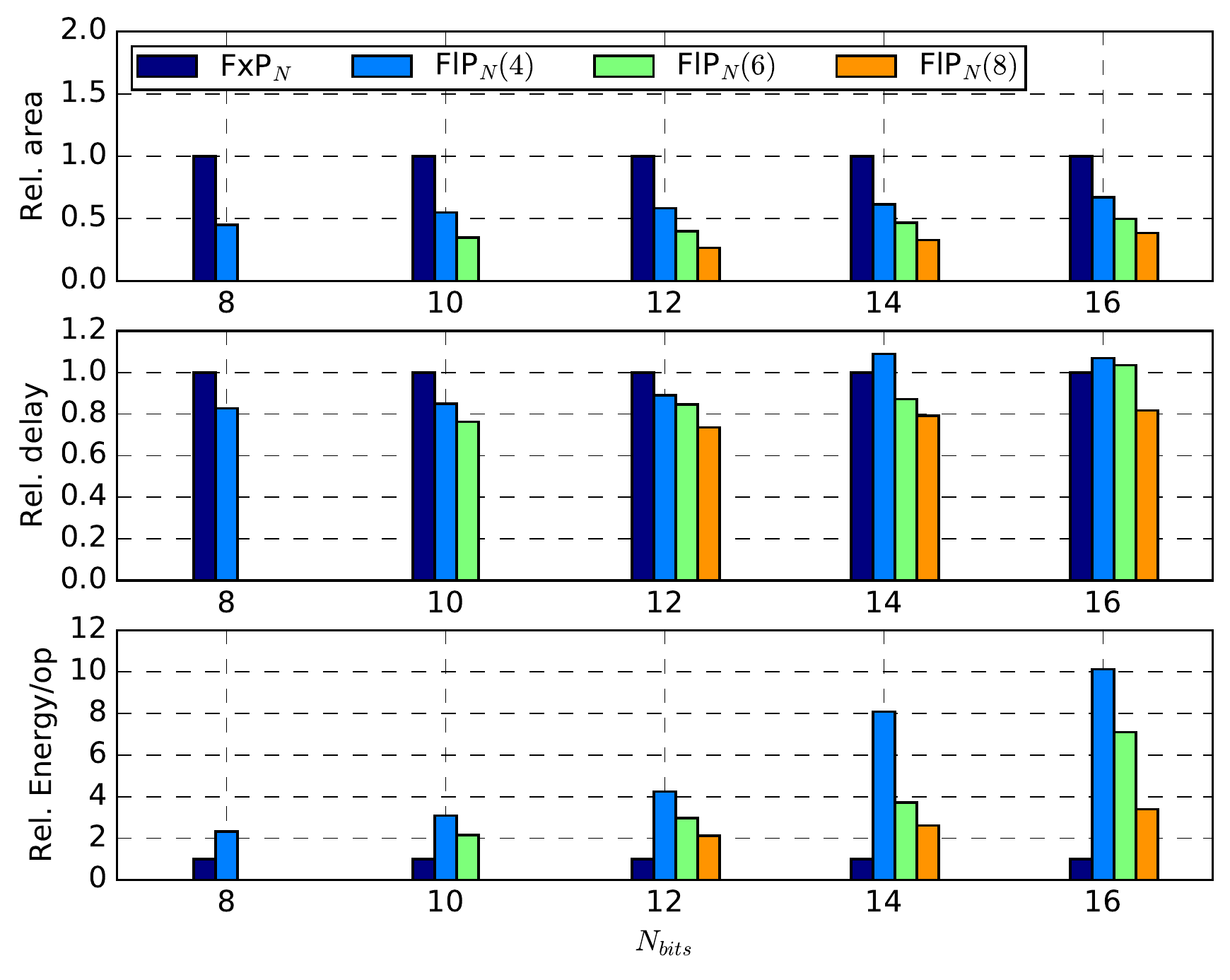}%
\caption{Relative area, delay and energy-per-operation between fixed-point and floating-point of multipliers for different bit-widths}
\label{fig:fxp_vs_FlP_mul}
\end{center}
\end{figure}
Figure~\ref{fig:fxp_vs_FlP_add} (resp. Figure~\ref{fig:fxp_vs_FlP_mul}) shows the area, delay and energy of adders (resp. multipliers) for different bit-widths, relative to the corresponding fixed-point operator. $\text{FlP}_{N}(E)$ represents $N$-bit floating-point with $E$ bits for exponent.
As discussed before in this chapter, the floating-point adder shows an important overhead compared to fixed-point. For any configuration, area and delay are around $3\times$ higher for floating-point. As a consequence, the higher complexity of the floating-point adder leads to $5\times$ to $12\times$ more energy per operation.

Results for the multipliers are very different. Floating-point multipliers are $2$-$3\times$ smaller than fixed-point. The control part of floating-point multiplier being less complex than for adder and, as multiplication is performed only on the mantissa, the area gets smaller. Timing is also slightly better for floating-point, but still constrained by operand shifts during computations. These shifts also significantly impact energy per operation, especially for large mantissas, which results in an overhead of $2\times$ to $10\times$ on the energy per operation for floating-point multiplication.

However, it must be kept in mind that, when using fixed-point numbers in an application, shifting is often needed at many steps during execution to align the number formats. The cost of shifting in the case of FxP is not considered in the results presented here, whereas it is already present in the case of floating-point. Thus, the advantage of fixed-point highlighted by Figures~\ref{fig:fxp_vs_FlP_add} and~\ref{fig:fxp_vs_FlP_mul} is expected to be tempered when full applications ae considered. This is the main objective of the next section.

\subsection{Application-level comparison}
\label{section:fxp_vs_flt_app}

In this section, floating-point and fixed-point operators are compared in the context of their use in applications. Indeed, as stated below, they have very different error nature and thus their error can not be fairly compared when considering only a single operation.
Both number representations are compared first using the K-Means clustering algorithm (also in~\cite{barrois:hal-01633723}) and then on the {Fast Fourier Transform (FFT)}.

\subsubsection{Results on K-Means clustering}
\label{subsection:fxp_vs_flt_kmeans}

This section first describes the K-means clustering algorithm before to provide comparative results between {FxP} and {FlP}. 

\begin{figure}[ht]
\begin{center}
\includegraphics[width=0.65\columnwidth]{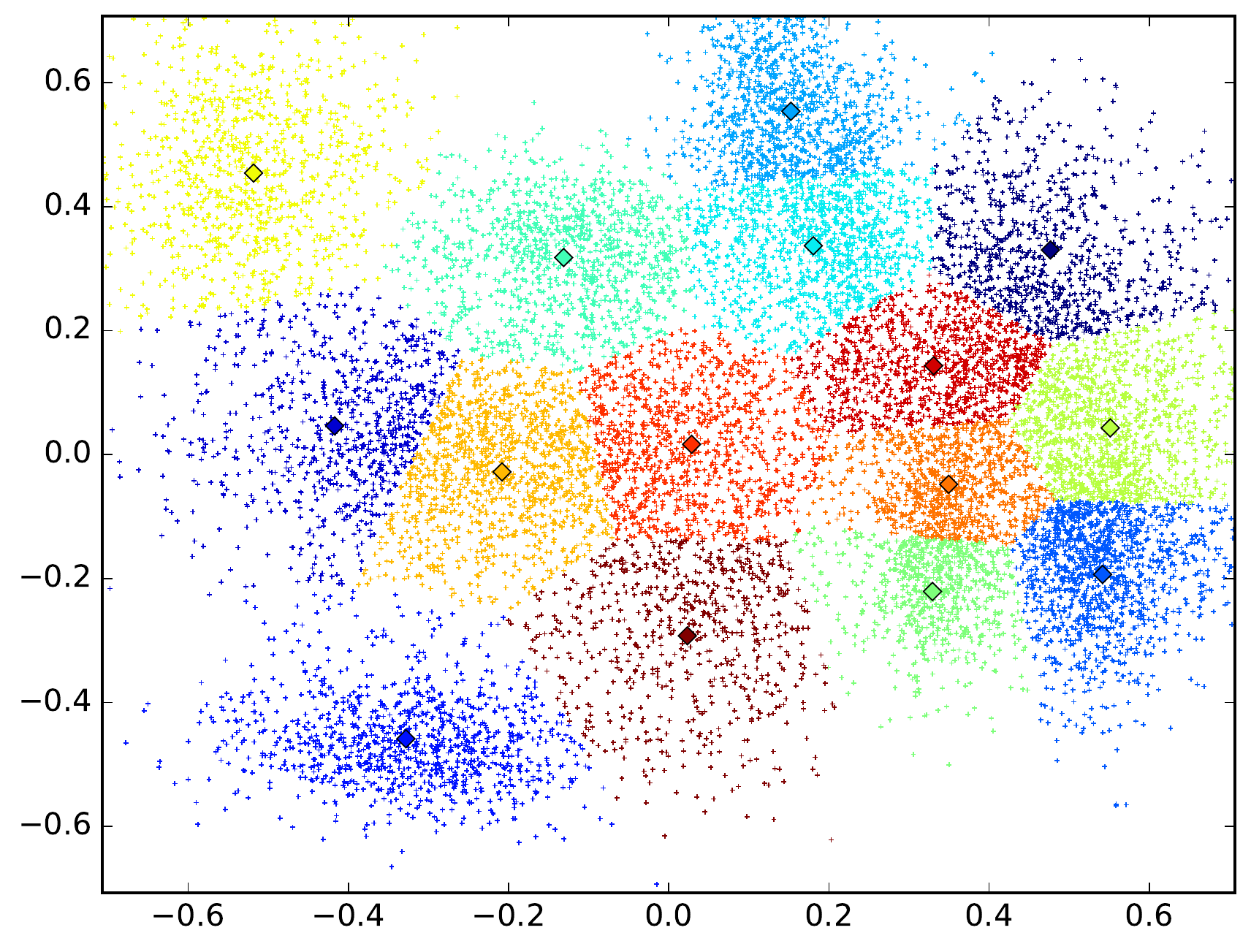}%
\caption{2-D K-Means clustering golden output example, obtained using double-precision (64-bit) floating-point}
\label{fig:kmeans_example}
\end{center}
\end{figure}

\paragraph{\textbf{\textit{K-Means clustering principle, algorithm and experimental setup}}}
\label{subsubsection:k_means_principle}
K-means clustering is a well-known method for vector quantization, which is mainly used in data mining, image classification or voice identification. It consists in organizing a multidimensional space into a given number of clusters, each being totally defined by its centroid. A given vector in the space belongs to the cluster in which it is nearest from the centroid. The clustering is optimal when the sum of the distances of all points to the centroids of the cluster they belong to, is minimal, which corresponds to finding the set of clusters $S = \left\{S_i\right\}_{i \in [0,k-1]}$ satisfying
\begin{equation}
   \underset{S}{arg\,min} \sum_{i=1}^{k} \sum_{x \in S_i}^{~} \left\| x - \mu_i \right\|^2 ,
\label{eq:Kmeans_equation}
\end{equation}
where $\mu_i$ is the centroid of cluster $S_i$.
Finding the optimal centroids position of a vector set is NP-hard. However, some iterative algorithms find good approximations of the optimal centroids by an \textit{estimation-maximization} process, with a linear complexity (linear with the number of clusters, number of data to process, number of dimensions, and number of iterations).
Lloyd's iterative algorithm~\cite{LLoydIT1982} is used in our case study. It is applied to bidimensional sets of vectors to ease display and interpretation of the results. From now, we only refer to Lloyd's algorithm in two dimensions. Figure~\ref{fig:kmeans_example} shows results of K-Means on a random set of input vectors, obtained using double-precision floating-point computations with a very restrictive stopping condition; these values are then considered as the reference golden outputs.

The algorithm consists of three main steps:
\begin{enumerate}
\item Initialization of the centroids.
\item Data labelling.
\item Centroid position update.
\end{enumerate}
Steps 2 and 3 are iterated until a stopping condition is met. In our case, the main stopping condition is when the difference of the sums of all distances from data points to their cluster's centroid between two iterations is less than a given threshold. A second stopping condition is the maximum number of iterations, required to avoid the algorithm getting stuck when the computations are too approximated to converge. The detailed algorithm for one dimension is given by Algorithm~\ref{algo:lloyd_k_means}. Inputs are represented by the vector $data$ of size $N_{data}$, output centroids by the vector $c$ of size $k$. The accuracy target for stopping condition is defined by $acc\_target$ and the maximum allowed number of iterations by $max\_iter$.
\begin{algorithm}
\caption{K-Means Clustering algorithm in one dimension}
\begin{algorithmic}
\Require $k \leq N_{data}$
\State $err \leftarrow +\infty$
\State $cpt \leftarrow 0$
\State $c \leftarrow init\_centroids (data)$
\Do\Comment{Main loop}
    \State $old\_err \leftarrow err$
    \State $err \leftarrow 0$
    \State $c\_tmp[0:k-1] \leftarrow 0$
    \State $min\_distance \leftarrow +\infty$
    \For{$d \in \{0:N_{data}-1\}$}
        \State $min\_distance \leftarrow +\infty$
        \For{$i \in \{0:k-1\}$} \Comment{Data labelling}
            \State $distance \leftarrow distance\_comp(data[d], c[i])$
            \If{$distance < min\_distance$}
                \State $min\_distance \leftarrow distance$
                \State $labels[d] \leftarrow i$
            \EndIf
        \EndFor
        \State $c\_tmp[labels[d]] \leftarrow c\_tmp[labels[d]] + data[d]$
        \State $counts[labels[d]] \leftarrow counts[labels[d]] + 1$
        \State $err \leftarrow err + min\_distance$
    \EndFor
    \For{$i \in \{0:k-1\}$} \Comment{Centroids position update}
        \If{$counts[i] \neq 0$}
            \State $c[i] \leftarrow c\_tmp[i]/counts[i]$
        \Else
            \State $c[i] \leftarrow c\_tmp[i]$
        \EndIf
    \EndFor
    \State $cpt \leftarrow cpt + 1$
\doWhile{$\left(\left|err-old\_err\right| > acc\_target\right) \lor \left(cpt < max\_iter\right)$}    
\end{algorithmic}
\label{algo:lloyd_k_means}
\end{algorithm}
In our study, we use several values for $acc\_target$, and $max\_iter$ is set to 150, which is never reached in practice. 

The impact of fixed-point and floating-point arithmetic on performance and accuracy is evaluated considering the distance computation function $distance\_comp$, defined by
\begin{equation}
    d \leftarrow (x-y) \times (x-y).
\label{eq:distance_comp}
\end{equation}
In the 2D case, the distance computation becomes
\begin{equation}
    d \leftarrow (x_0 - y_0) \times (x_0 - y_0) + (x_1 - y_1) \times (x_1 - y_1),
\label{eq:dist_equation2D}
\end{equation}
which is equivalent to 1~addition, 2 subtractions, and 2 multiplications. However, as distance computation is cumulative on each dimension, the hardware implementation relies only on 1~adder (accumulation), 1 subtracter, and 1 multiplier. \\

The experimental setup is divided into two parts: accuracy evaluation and cost/performance/energy estimation.
Accuracy estimation is performed on 20 data sets composed of $15 .10^{3}$ bidimensional data samples, all generated in a square delimited by  $\{\pm \sqrt{2}, \pm \sqrt{2}\}$, using Gaussian distributions with random covariance matrices around 15 random mean (centroid) points. Several accuracy targets are used to set the stopping condition: $10^{-2}$, $10^{-3}$, $10^{-4}$. As stated, the reference for accuracy estimation is IEEE-754 double-precision floating-point (Fig.~\ref{fig:kmeans_example}). The error metrics for the accuracy estimation are:
(i) the Mean Square Error of the resulting cluster Centroids (CMSE), and
(ii) the classification Error Rate (ER), which is defined as the proportion of points not being tagged by the right cluster identifier.
The lower the {CMSE}, the better the estimated position of centroids compared to golden output. Energy estimation is performed using the first of these 20 data sets, limited to $20.10^{3}$ iterations of distance computation for time and memory purposes. As data sets were generated around 15 points, the number of clusters researched is also set to 15. Area, latency of execution and energy are estimated using the same library and tools as in the previous section. Iterative distance computation is specified in C++ and HLS is used to generate the hardware under evaluation.

\paragraph{\textbf{\textit{Experimental results on K-Means clustering}}}
\label{subsubsection:kmeans_results}
Section~\ref{section:fxp_vs_flt_standalone} showed that FxP additions and multiplications consume less energy than their FlP counterparts for the same bit-width. However, these results do not yet consider the impact of the number formats on accuracy. This section details the impact of accuracy on the 2D K-means clustering algorithm.

A first qualitative study on the K-Means clustering showed that, to get correct results (no artefacts), FlP data must have a minimal exponent width of 5 bits in distance computation (smaller exponents are too inaccurate in low distance computations) and fixed-point data a minimal number of 3 bits for its integer part. Thus, all the following results use these two configurations and vary the mantissa and fractional part for FlP and FxP, respectively.
The total energy is defined as
\begin{equation}
    E_{\text{K-means}} = E_{\text{dc}} \times \left(N_{\text{it}} + N_{\text{cycles}} - 1 \right) \times N_{\text{data}},
\end{equation}
where $E_{\text{dc}}$ is the energy per distance computation estimated as in the previous section, $N_{\text{it}}$ the average number of iterations necessary to reach K-means stopping condition, $N_{\text{cycles}}$ the number of pipeline stages in the distance computation core, as determined by {HLS}, and $N_{\text{data}}$ the number of processed data per iteration.

\begin{figure}[ht]
\begin{center}
\subfloat[ct\_float$_{8}(5)$]{
\includegraphics[width=0.45\columnwidth]{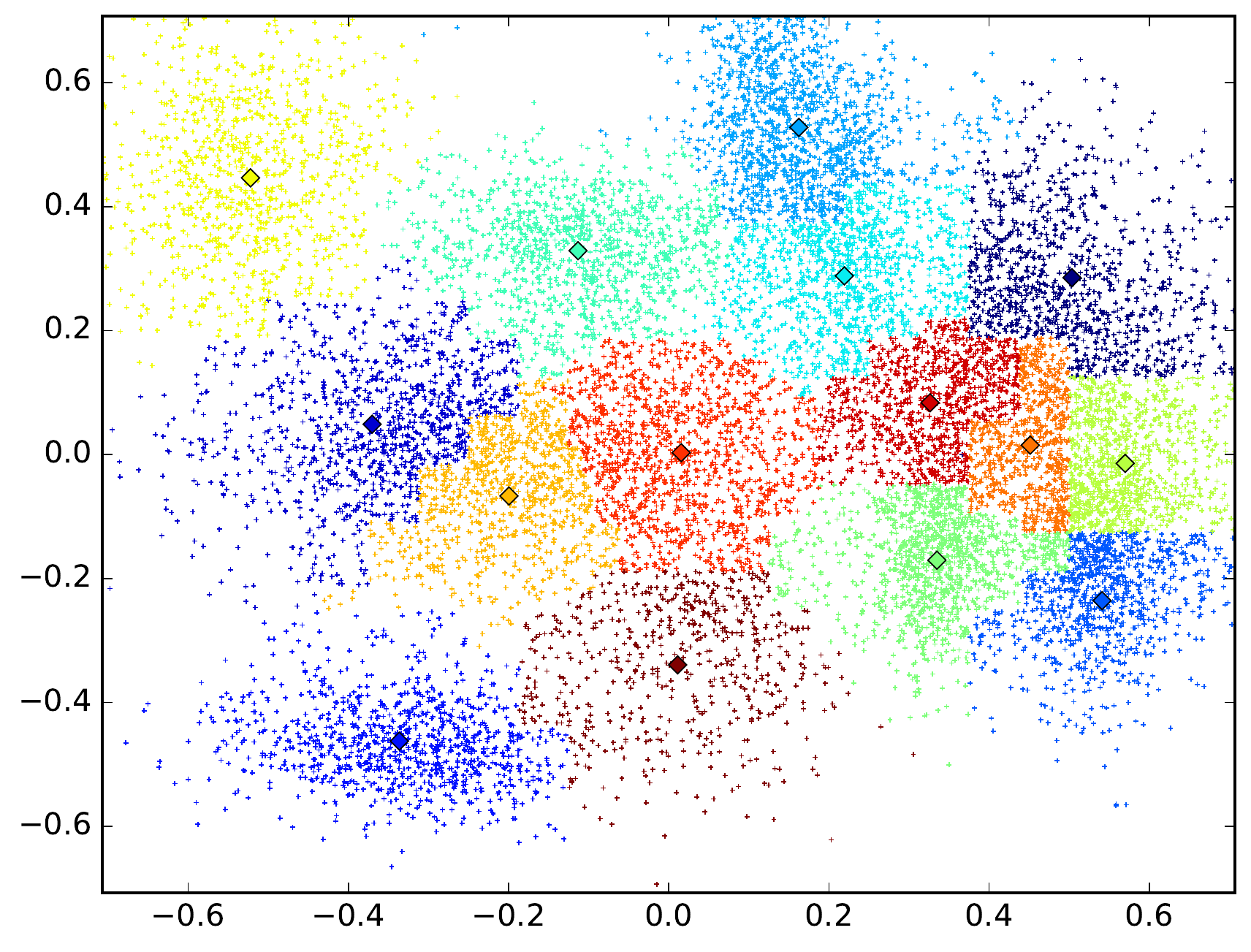}%
\label{subfig:ct_float_8_5}
}
\subfloat[ac\_fixed$_{8}(3)$]{
\includegraphics[width=0.45\columnwidth]{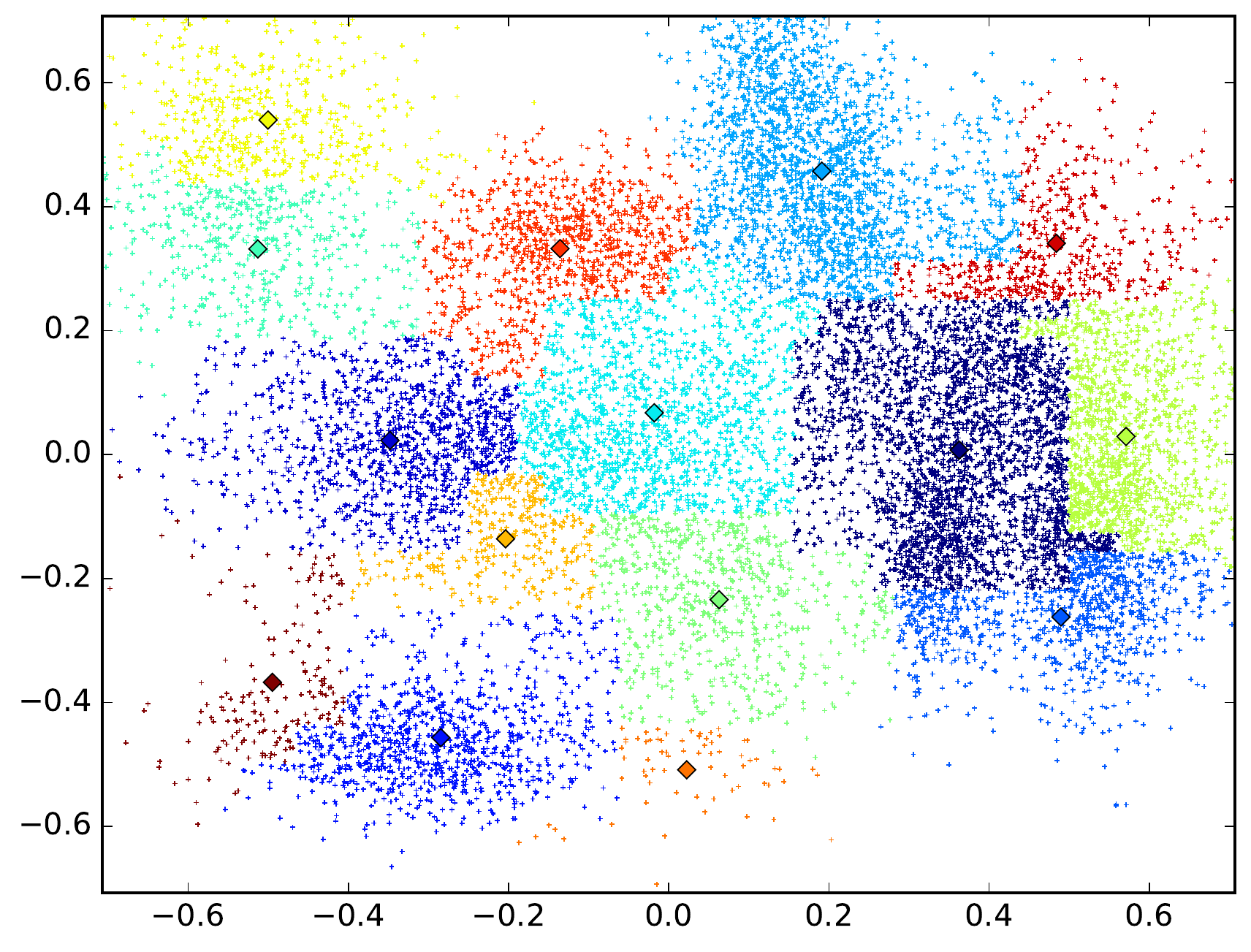}%
\label{subfig:ac_fixed_8_3}
}\\
\subfloat[ct\_float$_{16}(5)$]{
\includegraphics[width=0.45\columnwidth]{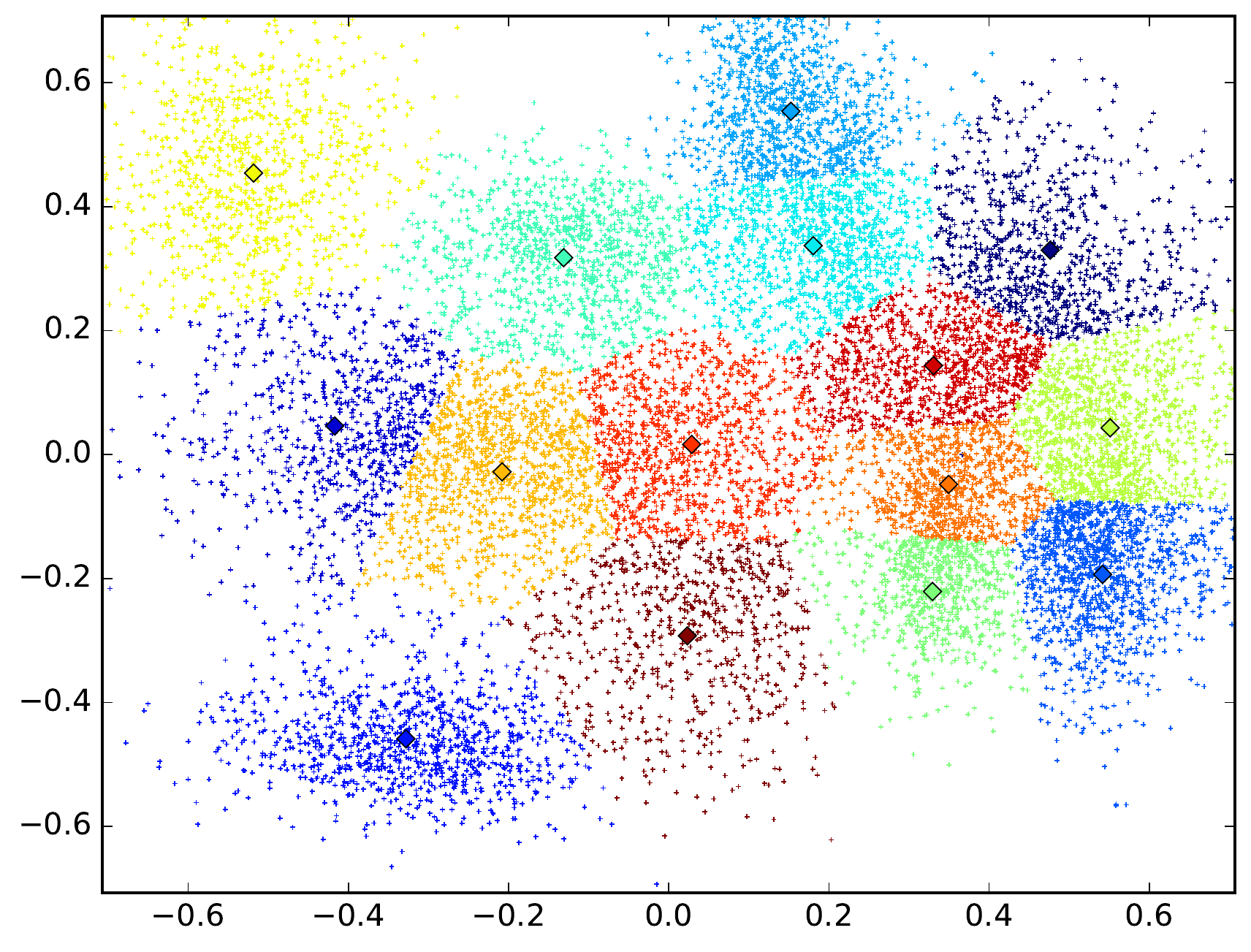}%
\label{subfig:ct_float_16_5}
}
\subfloat[ac\_fixed$_{16}(3)$]{
\includegraphics[width=0.45\columnwidth]{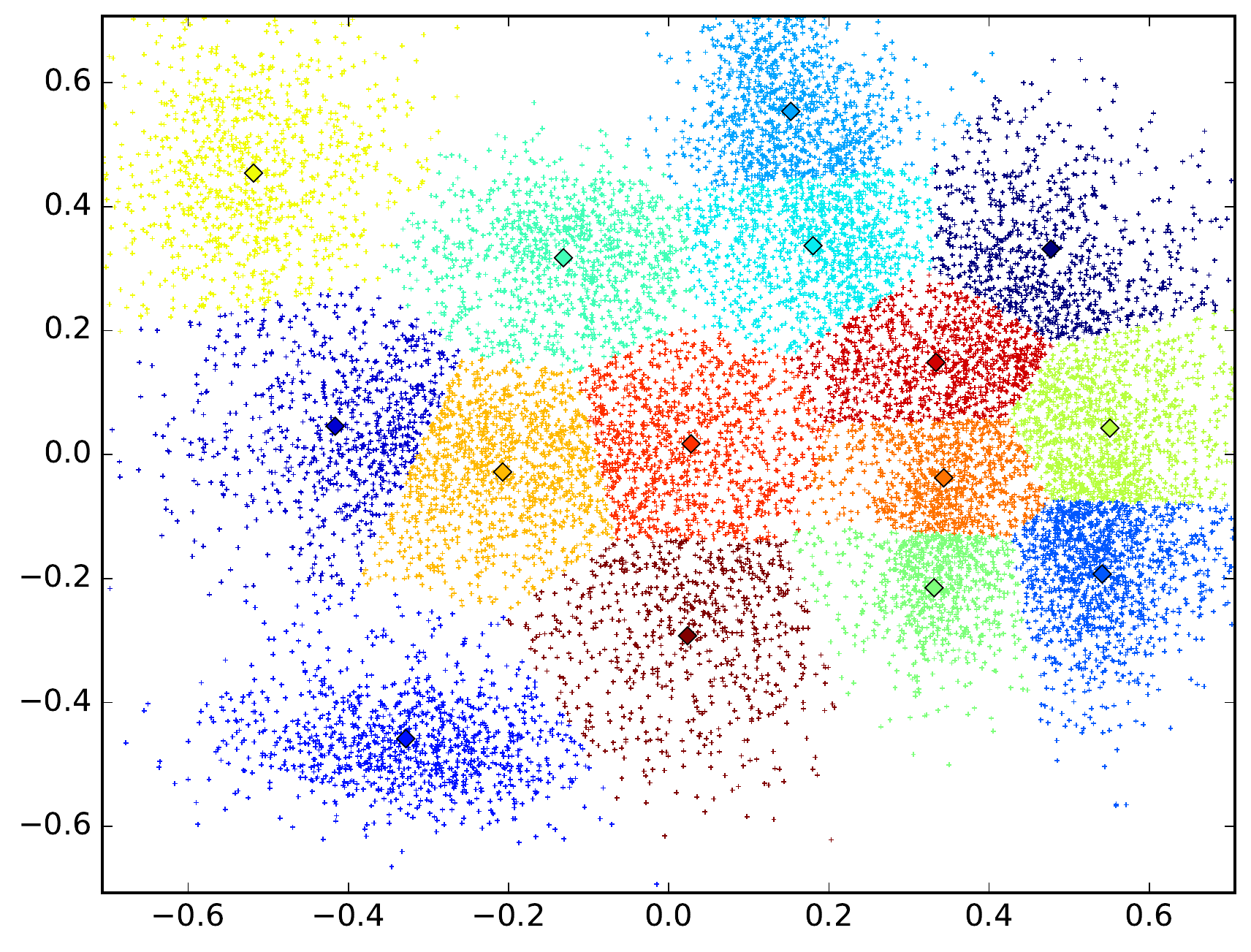}%
\label{subfig:ac_fixed_16_3}
}
\caption{K-Means clustering outputs for 8- and 16-bit floating-point and fixed-point, with accuracy target of $10^{-4}$}
\label{subfig:kmeans_outputs_flp_fxp}
\end{center}
\end{figure}

Results for 8-bit and 16-bit {FlP} and {FxP} arithmetic operators are detailed in Table~\ref{tab:8_16_bit_results}, with a stopping condition set to $10^{-4}$.
\begin{table}[htbp]
\centering
\begin{tabular}{|c|c|c|c|c|}
\hline
~&ct\_float$_{8}(5)$&ct\_float$_{16}(5)$&ac\_fixed$_{8}(3)$&ac\_fixed$_{16}(3)$\\
\hline
\hline
Area ($\mu m^2$)& 392.3	& 1148 & 180.7 & 575.1\\
\hline
$N_{\text{cycles}}$& 3 & 3 & 2 & 2\\
\hline
$E_{\text{dc}}$ ($nJ$) & $1.23\mathrm{E}{-4}$ & $5.99\mathrm{E}{-4}$ & $5.03\mathrm{E}{-5}$ & $3.25\mathrm{E}{-4}$\\
\hline
$N_{\text{it}}$& 8.35 & 59.3 & 14.9 & 65.1\\
\hline
$E_{\text{K-means}}$ ($nJ$)& 38.24 & 1100 & 23.90 & 644.34\\
\hline
CMSE & $1.75\mathrm{E}{-3}$ & $3.03\mathrm{E}{-7}$ & $1.85\mathrm{E}{-2}$ & $3.28\mathrm{E}{-7}$\\
\hline
Error Rate & 35.1 \% & 2.94 \% & 62.3 \% & 0.643 \%\\
\hline
\end{tabular}
\caption{8- and 16-bit area, energy and accuracy for K-Means clustering experiment}
\label{tab:8_16_bit_results}
\end{table}
For the 8-bit version of the algorithm, several interesting results can be highlighted. First, the custom FlP version is $2\times$ larger than FxP version, and FlP distance computation consumes $2.44\times$ more energy than FxP. However, the FlP version of K-means converges in $8.35$ cycles on average, against $14.9$ cycles for FxP. This results in making the floating-point version for the whole K-means algorithm consuming only $1.6\times$ more energy than fixed-point. Moreover, the FlP version provides a huge advantage in terms of accuracy of results. Indeed, {CMSE} is $10\times$ better for FlP and {ER} is $1.8\times$ better. Figures~\ref{subfig:ct_float_8_5} and~\ref{subfig:ac_fixed_8_3} show the output for floating-point and fixed-point 8-bit computations, applied on the same inputs as the golden output of Fig.~\ref{fig:kmeans_example}. A very neat stair-effect on data labelling is clearly visible, which is due to the high quantization levels of the 8-bit representation. However, in the floating-point version, the positions of clusters' centroid is very similar to the reference, which is not the case for fixed-point.

For the 16-bit version, all results are in favor of fixed-point, floating-point being twice bigger and consuming $1.7\times$ more energy. FxP also provides slightly better error results (2.9\% for {ER} vs. 0.6\%). Figures~\ref{subfig:ct_float_16_5} and~\ref{subfig:ac_fixed_16_3} show output results for 16-bit floating-point and fixed-point. Both are very similar and nearly equivalent to the reference, which reflects the high success rate of clustering.

\begin{figure}[hp]
\begin{center}
\includegraphics[width=0.7\columnwidth]{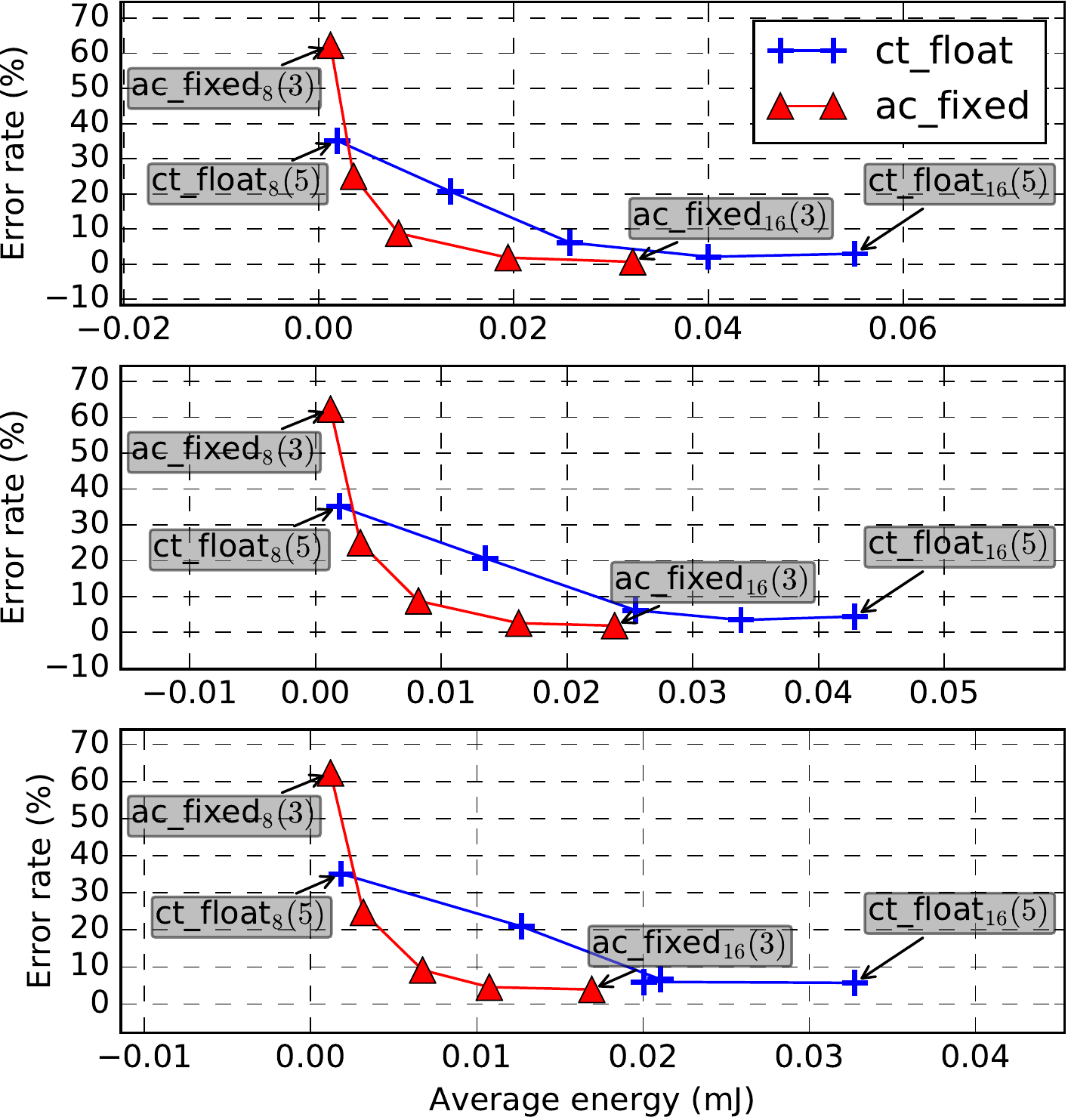}%
\caption{Energy vs. classification error rate for K-Means clustering with stopping conditions of $10^{-4}$ (top), $10^{-3}$ (center) and $10^{-2}$ (bottom)}
\label{fig:kmeans_energy_vs_er}
\end{center}
\end{figure}

The competitiveness of {FlP} over {FxP} on small bit-widths, and the higher efficiency of {FxP} on larger bit-widths, are confirmed by Fig.~\ref{fig:kmeans_energy_vs_er} depicting energy vs. classification error rate. Indeed, for different accuracy targets ($10^{-\{2,3,4\}}$), only 8-bit FlP provides higher accuracy for a comparable energy cost, whereas 10- to 16-bit FxP versions reach an accuracy equivalent to FlP with much less energy. The stopping condition does not seem to have a major impact on the relative performance.

\subsubsection{Results on Fast Fourier Transform}
\label{subsection:fxp_vs_flt_FFT}
In the previous section, a comparative study between custom FxP and FlP was performed on K-means, showing that, contrary to what could be expected, floating-point can be very competitive for small bit-widths.
In this section, a similar study is performed on the Fast Fourier Transform ({FFT}).

The implementation of the {FFT} is Radix-2 Decimation-in-Time (DIT), which is the most common form of the Cooley-Tukey algorithm~\cite{CooleyMC1965}. For the hardware estimation, only the kernel computations of the {FFT} are considered, i.e., 
\begin{eqnarray}
\label{eq:FFT_DIT}
X_{k} = E_{k} + e^{-\frac{2 \pi i}{N} k} O_{k}, \nonumber \\
X_{k+\frac{N}{2}} = E_{k} - e^{-\frac{2 \pi i}{N} k} O_{k},
\end{eqnarray}
equivalent to 6 additions/subtractions and 4 multiplications. For each version of the {FFT}, all constants and variables are represented with the same parameters (same bit-width, same integer part width for {FxP}, same exponent width for {FlP}). The absence of over/underflow for the {FxP} version is ensured. For the {FlP} version, the repartition of the exponent and mantissa widths is chosen for giving the smallest error after an exhaustive search. For hardware performance estimation, only FFT-16 (FFT on $N=16$ samples) was characterized. 
The error metric is the Mean Square Error ({MSE}) at the output compared to double-precision floating-point.

\begin{figure}[htbp]
\begin{center}
\includegraphics[width=0.75\columnwidth]{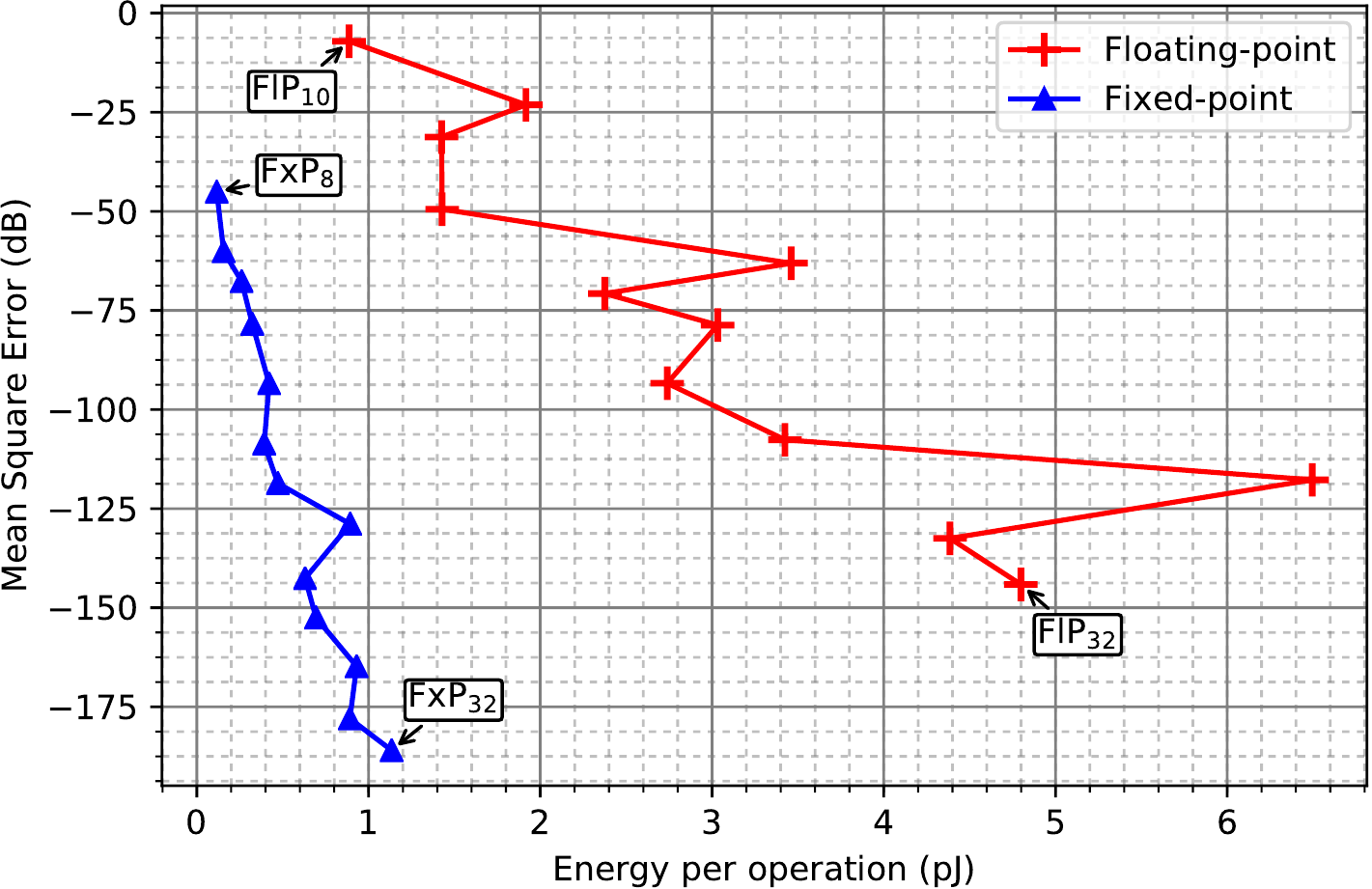}
\caption{Fixed-point and floating-point energy per operation (pJ) vs {MSE} for FFT-16 for different bit-widths}
\label{fig:FFT_flp_vs_fxp}
\end{center}
\end{figure}
Energy per operation (pJ) related to error (MSE in dB) for {FFT}-16 is depicted in Fig.~\ref{fig:FFT_flp_vs_fxp}. The error-energy trade-off is better when reaching the bottom-left corner. For each curve, each point going from the top left to the bottom right represents an increase of two digits in the bit-width.

For this application, the advantage is clearly in favour of fixed-point. Indeed, for any identical bit-width, FxP outperforms FlP in both energy and accuracy. As already showed in Section~\ref{section:fxp_vs_flt_standalone}, FlP operations, and additions in particular, are much more expensive than FxP. However, {FFT} output quality is not as dependent on accuracy on a dynamic as large as for K-means clustering. This makes FlP even less accurate than FxP at equal bit-width, because of a smaller significant part, mantissa for floating-point, all bits for fixed-point. Indeed, in the experiment, the exponent takes 7 bits of the total width, which are not assigned to more accuracy on the significant part.
Another interesting point is the data points presenting an \emph{energy peak}, which are occurring for 12-, 18- and 28-bit floating-point and 22-bit fixed-point. These peaks are most probably due to differences of implementation in the {HLS} process. E.g, larger adder or multiplier structures may have been selected by the tool to meet constraints of delays, leading to energy overhead.

\section{Conclusion}

Computing with low-precision arithmetic is an efficient way to maximize performance-per-Watt~\cite{barrois:hal-01423147}. Customization is then ruled by finding a trade-off between reducing precision to improve energy efficiency and respecting constraints on application output quality. This chapter mainly focused on floating-point and fixed-point number representations, presented their principle and some opportunities for arithmetic customization on both formats, and provided a comparison between their cost, performance and energy, as well as their impact on accuracy during computations.  

Comparing floating-point and fixed-point arithmetic at the operator level gives a clear advantage in area, delay and energy efficiency for fixed-point.
However, when considering real applications, e.g., the study on K-means clustering algorithm in this chapter, custom floating-point arithmetic can provide interesting features, and tends to show a better energy/accuracy trade-off for very small bit-widths (8 bits in this study). However, the advantage come back to fixed-point when the considered application is an FFT (same is true for digital filters or most classical signal and image processing algorithms).
One explanation is that applications requiring both large dynamic range and high accuracy, are more tolerant to low precision when the floating-point representation is used. An interesting follow-up of this study would be to consider larger FFT, which would lead to larger dynamic range, and to see how the MSE vs. energy-per-operation would scale for both representations.

Another important aspect of the study is from a hardware-design point of view. Floating-point is very complex compared to fixed-point arithmetic for large bit-widths, but the overhead is shrinking when lowering precision. Moreover, the cost of multiplication can be considered at the advantage of floats. Also, the overhead of scaling instructions required to be added to deal with fixed-point data types is often not studied. It is an interesting perspective to include this overhead in the choice of the right representation.

Hence, in the aim of designing general-purpose low-energy processors, low-precision floating-point arithmetic can provide major advantages compared to classical integer operators embedded in microcontrollers, with a better compromise between ease of programming, energy efficiency and computing accuracy. In the context of inference and training of deep neural networks, custom float is also a serious competitor. This is one of the objectives of Chapter 15 presenting opportunities for Approximations in Deep Learning.

\bibliographystyle{ieeetr}
\bibliography{reference}

\end{document}